\documentclass[aps,prd,twocolumn,superscriptaddress,nofootinbib,groupedaddress]{revtex4-1}

\pdfoutput=1

\usepackage{slashed}
\usepackage{color, verbatim}
\usepackage{latexsym}
\usepackage{amssymb}
\usepackage{amsmath}
\usepackage{graphicx}\graphicspath{{figs/}}
\usepackage{arydshln}
\usepackage[dvipsnames]{xcolor}
\usepackage{adjustbox}
\usepackage{easybmat}
\usepackage{bbm}
\usepackage{bm}
\usepackage{adjustbox}
\usepackage{booktabs}
\usepackage{multirow}
\usepackage{rotating}
\usepackage{soul}
\usepackage{hyperref}
\usepackage{mathtools}

\allowdisplaybreaks

\begin{document}

\title{Positivity bounds in the Standard Model effective field theory beyond tree level}

\author{Mikael Chala}
\email{mikael.chala@ugr.es}
\author{Jose Santiago}
\email{jsantiago@ugr.es}
\affiliation{Departamento de F\'isica Te\'orica y del Cosmos,
Universidad de Granada, E--18071 Granada, Spain}

\begin{abstract}
Focusing on four-Higgs interactions, we analyse the robustness of tree-level-derived positivity bounds on Standard Model effective field theory (SMEFT) operators under quantum corrections. Among other results, we demonstrate that: \textit{(i)} Even in the simplest extensions of the Standard Model, \textit{e.g.} with one new scalar singlet or with a neutral triplet, some positivity bounds are strictly violated; \textit{(ii)} the mixing of the dimension-eight operators under renormalisation, which we compute here for the first time, can drive them out of their positivity region; \textit{(iii)} the running of the dimension-eight interactions triggered by solely dimension-six terms respects the positivity bounds. Our results suggest, on one hand, that departures from positivity within the SMEFT, if ever found in the data, do not necessarily imply the breaking of unitarity or causality, nor the presence of new light degrees of freedom. On the other hand, they lead to strong constraints on the form of certain anomalous dimensions.
\end{abstract}

\maketitle

\newpage

\section{Introduction}
\label{sec:intro}
Effective-field theories (EFT) are the right tool to describe particle physics in the presence of significant mass gaps. In particular, the Standard Model EFT (SMEFT), see Ref.~\cite{Brivio:2017vri} for a review, seems to be the most appropriate theory at energies $100\, \text{GeV}\lesssim E\lesssim \text{TeV}$, given that no new resonances have been found in this regime. 

The parameters of the SMEFT Lagrangian have been subject of experimental scrutiny for many years. By now, many directions in the SMEFT, although certainly not all, have been severely constrained; see for example Refs.~\cite{deBlas:2016nqo,Ellis:2020unq,Ethier:2021bye}. 

More recently, though, there has been a huge progress in narrowing the SMEFT landscape purely from theoretical arguments~\cite{Zhang:2018shp,Bi:2019phv,Remmen:2019cyz,Gu:2020ldn,Bonnefoy:2020yee,Remmen:2020vts,Zhang:2020jyn,Gu:2020thj,Fuks:2020ujk,Yamashita:2020gtt,Remmen:2020uze}. These rely on the basic principles of quantum mechanics and relativity, and in particular on the analyticity and unitarity of the $S$-matrix~\cite{Adams:2006sv}. The corresponding bounds appear in the form of constraints on the sign of certain combinations of Wilson coefficients, and they are commonly known as \textit{positivity bounds}. Often, they are complementary to current experimental limits~\cite{Remmen:2019cyz}.

Positivity bounds have been studied mostly under the assumption that all operators involved arise at the same order in perturbation theory. There is also little knowledge on the evolution of these constraints under renormalisation group equations (RGE)~\cite{Bellazzini:2020cot,Arkani-Hamed:2020blm}. In this paper, we examine these aspects more closely, focusing on the dimension-eight operators with four Higgs fields. A nice property of these operators is that, contrary to, for example, other quartic gauge boson couplings which have been matter of study within the context of positivity, they can arise at tree level in ultraviolet (UV) completions of the SM~\cite{Craig:2019wmo}.

The paper is organised as follows. In Section~\ref{sec:tree}, we review briefly the derivation of positivity bounds at tree level, while we dedicate Section~\ref{sec:loop} to analysing important differences that arise at one loop.  We substantiate this discussion with explicit examples of matching of UV models onto the dimension-eight SMEFT in Section~\ref{sec:matching}, as well as with the  computation of the RGEs of the four-Higgs interactions in Section~\ref{sec:running}. We conclude in Section~\ref{sec:conclusions}. Appendix~\ref{app:details} is dedicated to highlighting some technical details of the calculations.

\section{Positivity bounds at tree level}
\label{sec:tree}
To fix notation, let us first write the SM Lagrangian:
\begin{align}
 \mathcal{L}_\text{SM} &= -\frac{1}{4}G_{\mu\nu}^A G^{A\mu\nu}-\frac{1}{4}W_{\mu\nu}^I W^{\mu\nu I}-\frac{1}{4}B_{\mu\nu}B^{\mu\nu}\nonumber\\
 &+(D_\mu H)^\dagger (D^\mu H) + \mu^2 H^\dagger H-\lambda (H^\dagger H)^2\nonumber\\
 &+i (\overline{q}\slashed{D}q+\overline{u}\slashed{D}u+\overline{d}\slashed{D}d+\overline{l}\slashed{D}l+\overline{e}\slashed{D}e)\nonumber\\
 &-(\overline{q} Y_d Hd + \overline{q}Y_u\tilde{H} u+\overline{l}Y_e He+\text{h.c.})\,.
\end{align}
We have introduced $l$ and $e$ for the left-handed (LH) and right-handed (RH) leptons, respectively; and $q$ and $u,d$ for the LH and RH quarks, respectively. $G$, $W$ and $B$ represent the gauge bosons of $SU(3)_c$, $SU(2)_L$ and $U(1)_Y$, and $H = \frac{1}{\sqrt{2}}(\phi_1+i \phi_2, \phi_3 + i\phi_4)^T$ stands for the Higgs doublet. We have also defined $\tilde{H}=\epsilon H$, with $\epsilon$ being the fully anti-symmetric tensor.

Our convention for the covariant derivative is:
\begin{equation}
 D_\mu = \partial_\mu -ig_1 Y B_\mu-ig_2\frac{\sigma^I}{2}W_\mu^I-ig_s\frac{\lambda^A}{2}G_\mu^A\,.
\end{equation}
$Y$ stands for the hypercharge, $g_1$, $g_2$ and $g_3$ represent the $U(1)_Y$, $SU(2)_L$ and $SU(3)_c$ gauge couplings; and $\sigma^I$ and $\lambda^A$ denote the Pauli and Gell-Mann matrices, respectively.

We disregard the Yukawa couplings throughout this paper, since they do not play any role in any of our discussions. Likewise, we work in the approximation $\mu^2\to 0$. All our results are then valid up to $\mu^2/\Lambda^2$ corrections, where $\Lambda$ represents the SMEFT cutoff.
Moreover, unless otherwise stated, we also assume $g_1,g_2,g_3\to 0$, as our points are made clearer within this approximation.
\begin{table}[t]
 \begin{tabular}{cc}
  \hline\hline
  $\mathcal{O}_{H^4 D^2}^{(1)}$ & $(H^\dagger H)\Box (H^\dagger H)$\\[0.1cm]
  $\mathcal{O}_{H^4 D^2}^{(2)}$ & $(H^\dagger D^\mu H)^* (H^\dagger D_\mu H)$\\[0.1cm]
  \hline
  $\mathcal{O}_{H^4 D^4}^{(1)}$ & $(D_\mu H^\dagger D_\nu H) (D^\nu H^\dagger D^\mu H)$\\[0.1cm]
  $\mathcal{O}_{H^4 D^4}^{(2)}$ & $(D_\mu H^\dagger D_\nu H) (D^\mu H^\dagger D^\nu H)$\\[0.1cm]
  $\mathcal{O}_{H^4 D^4}^{(3)}$ & $(D_\mu H^\dagger D^\mu H) (D^\nu H^\dagger D_\nu H)$\\
  \hline\hline
 \end{tabular}
 \caption{\it Independent four-Higgs operators at dimension six (top) and dimension eight (bottom).}\label{tab:operators}
\end{table}

The SMEFT extends the SM Lagrangian with operators of dimension higher than four, suppressed by increasing powers of the cutoff $\Lambda$. We neglect operators of dimension higher than eight as well as lepton- and baryon-number violating interactions. This leaves us with operators of dimension six and eight only, that we choose to describe using the physical
bases of interactions reported in Ref.~\cite{Grzadkowski:2010es} and Ref.~\cite{Murphy:2020rsh}, respectively.

Let us consider the process $\phi_i\phi_j\to \phi_i\phi_j$. At low energies, it can be described by four-Higgs interactions:
\begin{equation}\label{eq:lagintro}
 \mathcal{L} = \cdots - \lambda |H|^4+ \dfrac{c_{H^4 D^2}^{(i)}}{\Lambda^2} \mathcal{O}_{H^4 D^2}^{(i)} + \dfrac{c_{H^4 D^4}^{(j)}}{\Lambda^4} \mathcal{O}_{H^4 D^4}^{(j)}\,;
\end{equation}
see Table~\ref{tab:operators} for the definition of the operators.
The most common nomenclature for the dimension-six operators is $\mathcal{O}_{\phi\Box}$ (for $i=1$) and $\mathcal{O}_{\phi D}$ (for $i=2$)~\cite{Grzadkowski:2010es}; we find however convenient to work with the different naming for clarity of the exposition. 

\begin{figure}[t]
 \includegraphics[width=0.8\columnwidth]{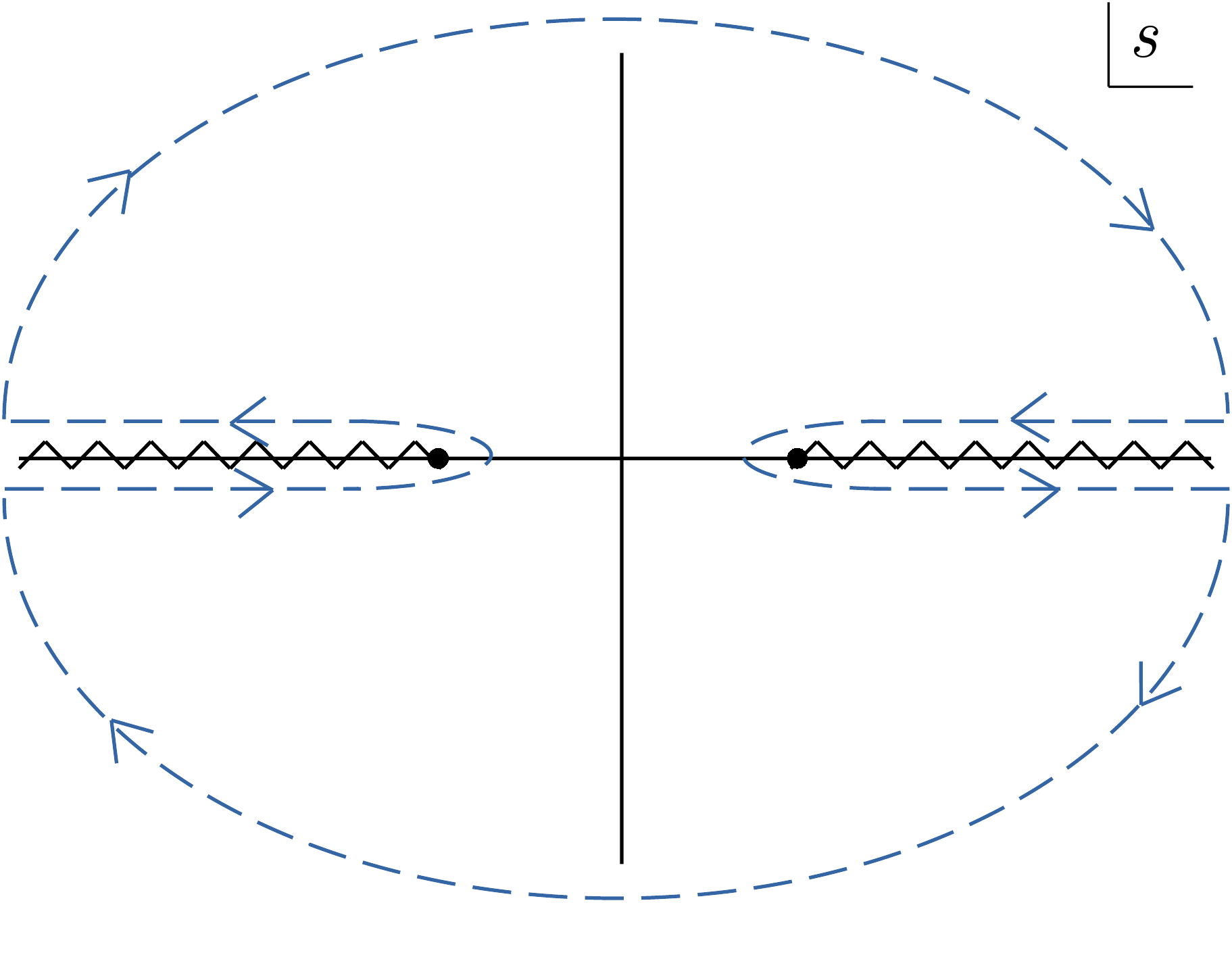}
 \caption{\it Structure of singularities of the forward two-to-two amplitude $\mathcal{A}(s)$ in the complex plane of $s$. We also show the contour of integration used in the derivation of positivity bounds.}\label{fig:complexplane}
\end{figure}

The usual derivation of bounds on the dimension-eight coefficient works as follows. First, we \textit{assume} that in the UV the forward scattering amplitude $\mathcal{A}(s)=\mathcal{A}(s,t=0)$ for the process of interest
is analytic in the complex plane with, at most, branch cuts in the real $s$ axis starting at $|s|>0$; see Fig.~\ref{fig:complexplane}. Note that this implicitly assumes that the running of the EFT Wilson coefficients, and therefore a branch cut all the way to the origin, is either absent or negligible. In this sense, we can talk about \textit{the} EFT Wilson coefficients, without mention to any renormalisation scale.

We then consider the integral $\mathcal{I} = \oint\mathcal{A}(s)/s^3$ around a small circular path enclosing $s=0$. By Cauchy's theorem, $\mathcal{I}$ is fixed by the residue of the integrand at the origin.

Now, the circular path can be deformed to an infinitely large contour as the one shown in the figure. The contribution from the circular sectors to $\mathcal{I}$ vanishes because the amplitude falls fast enough at infinity~\cite{Froissart:1961ux,Martin:1962rt}, while the contribution from the discontinuities can be related to the imaginary part of the forward amplitude~\cite{Adams:2006sv}, which by virtue of the optical theorem is positive.
Altogether, we obtain that the residue of $\mathcal{A}(s)/s^3$ at the origin is positive, or in other words: 
\begin{equation}\label{eq:positivity}
 \frac{d^2 \mathcal{A}(s)}{ds^2} \bigg\rvert_{s=0}> 0\,.
\end{equation}
This residue can be computed in the EFT. Using Eq.~\eqref{eq:lagintro} for the process $\phi_1\phi_2\to \phi_1\phi_2$ \textit{at tree level}, we find that
\begin{equation}
 \mathcal{A}(s) = -2\lambda  + c_{H^4 D^4}^{(2)} \frac{s^2}{\Lambda^4}\,,
\end{equation}
which gives
\begin{equation}\label{eq:firstconstraint}
 c_{H^4 D^4}^{(2)} > 0\,.
\end{equation}
The processes $\phi_1\phi_3\to \phi_1\phi_3$ and $\phi_1\phi_1\to \phi_1\phi_1$ imply the constraints:
\begin{align}\label{eq:otherbounds1}
 c_{H^4 D^4}^{(1)} + c_{H^4 D^4}^{(2)} > 0\,,\\\label{eq:otherbounds2}
 c_{H^4 D^4}^{(1)} + c_{H^4 D^4}^{(2)} +c_{H^4 D^4}^{(3)} > 0\,,
\end{align}
respectively;
see Ref.~\cite{Remmen:2019cyz}. Compatible bounds were also obtained in Ref.~\cite{Bi:2019phv} in the broken phase of the Higgs.

\section{Positivity bounds at one loop}

\label{sec:loop}
In order to analyse the fate of the above results when moving to one loop, we must first notice that the Wilson coefficients of the operators, at the EFT cutoff scale $\Lambda$, admit themselves a perturbative expansion. To make it clear, we introduce a weak coupling $g$ and write:
\begin{align}
 c_{H^4 D^4}^{(j)} &= g\, c_{H^4 D^4}^{(j)\, \text{tree}} + g^2\, c_{H^4 D^4}^{(j)\,\text{loop}} + \cdots\,,
\end{align}
and similarly for $c_{H^4 D^2}^{(i)}$ and $\lambda$. Thus, the forward amplitude for $\phi_1\phi_2\to\phi_1\phi_2$ scattering to order $\mathcal{O}(g^2)$ in a neighbourhood of $s=0$ reads:\\
%
%
\begin{align}\label{eq:amplitude}
\mathcal{A}(s) &\sim -2g \lambda^{\text{tree}}+g^2[-2\lambda^{\text{loop}}+\frac{3}{2\pi^2}(\lambda^{\text{tree}})^2\log{\frac{\Lambda^2}{s}}]\nonumber\\
 &+ (g\, c_{H^4 D^4}^{(2)\,\text{tree}} + g^2 c_{H^4 D^4}^{(2) \,\text{loop}} -\frac{\beta_{H^4 D^4}^{(2)}}{2}\log{\frac{\Lambda^2}{s}})\frac{s^2}{\Lambda^4}\,,
\end{align}
up to finite terms proportional to the tree level Wilson coefficients. The
 $\beta_{H^4D^4}^{(2)}$ stands for the $\beta$ function of $c_{H^4 D^4}^{(2)}$, defined by $\mu\, d c_{H^4D^4}^{(2)}/d\mu=\beta_{H^4 D^4}^{(2)}$. This function receives two contributions, corresponding to the renormalisation triggered by pairs of dimension-six interactions and to that driven by dimension-eight operators via $\lambda$. Schematically:
\begin{equation}
 \frac{1}{g^2}\beta_{H^4 D^4}^{(2)}\sim \gamma'_{ij} c_{H^4 D^2}^{(i)\,\text{tree}} c_{H^4 D^2}^{(j)\,\text{tree}} + \gamma_i\, \lambda^\text{tree} c_{H^4 D^4}^{(i)\,\text{tree}}\,.
\end{equation}
Several interesting conclusions can be derived from considering Eq.~\eqref{eq:amplitude} in different limits.

To start with, let us assume that none of the effective interactions is generated at tree level, and $\lambda^\text{tree}=0$ as well. Then, $\mathcal{A}(s)$ is simply:
\begin{align}
 \mathcal{A}(s) \sim c_{H^4 D^4}^{(2)\,\text{loop}}\frac{s^2}{\Lambda^4}\,.
\end{align}
Upon making the same reasoning as in Section~\ref{sec:tree} leading to Eq.~\eqref{eq:positivity}, we obtain $c_{H^4 D^4}^{(2),\,\text{loop}}>0$.

Let us now turn our attention to the case in which all effective operators but $\mathcal{O}_{H^4 D^4}^{(2)}$ can arise at tree level, and still $\lambda^\text{tree}=0$. The amplitude $\mathcal{A}(s)$ near $s=0$ is:
\begin{align}
 \mathcal{A}(s) \sim g^2 (c_{H^4 D^4}^{(2)\,\text{loop}} - \frac{\gamma'_{ij}}{2} c_{H^4 D^2}^{(i)\,\text{tree}} c_{H^4 D^2}^{(j)\,\text{tree}}\log{\frac{\Lambda^2}{s}})\frac{s^2}{\Lambda^4}\,.
\end{align}
In this case, the branch cut all the way to $s=0$ originated by the logarithm prevents using the argument outlined in Section~\ref{sec:tree}. To circumvent this obstacle, one can include a small mass $m$ for the Higgs, thus generating an analytic region around $s=0$. This amounts to deforming the logarithm $\log{\Lambda^2/s}\to\log{[\Lambda^2/(s+m^2)]}$. We can subsequently study the limit $m^2\to 0$ upon expanding the logarithm in powers of $s/m^2$; see Refs.~\cite{Adams:2006sv,Arkani-Hamed:2020blm}. In such limit, the dominant contribution is:
\begin{equation}
 \mathcal{A}(s) \sim -g^2 \frac{\gamma'_{ij}}{2}  \log{\frac{\Lambda^2}{m^2}} c_{H^4 D^2}^{(i)\,\text{tree}} c_{H^4 D^2}^{(j)\,\text{tree}}\frac{s^2}{\Lambda^4} + \mathcal{O}(s^3)\,.
\end{equation}
A first implication of this result is that $c_{H^4 D^4}^{(2)\,\text{loop}}$ can have either sign without affecting the positivity of the forward amplitude. Therefore, the bound obtained in Section~\ref{sec:tree}, $c_{H^4 D^4}^{(2)}>0$, does not necessarily hold in models in which this coefficient arises only in loops, provided that other operators are generated at tree level.

Further, requiring the second derivative of $\mathcal{A}(s)$ to be positive at the origin implies very severe constraints on the running of $c_{H^4 D^4}^{(2)}$ triggered by pairs of dimension-six interactions, namely:
\begin{equation}\label{eq:dim6ren}
 \gamma'_{ij} c_{H^4 D^2}^{(i)\,\text{tree}} c_{H^4 D^2}^{(j)\,\text{tree}} < 0\,.
\end{equation}
Note that, because arbitrary values of the dimension-six Wilson coefficients are compatible with the  assumption $c_{H^4 D^4}^{(2)\,\text{tree}}=0$~\cite{Chala:2021pll}, and given that $\lambda$ can be always made zero by just tuning the renormalisable Lagrangian, the bound above is completely general. Moreover, using the exact same reasoning one concludes that this inequality is valid even when including fermionic dimension-six operators, the relevant of which are:
\begin{align}
\mathcal{O}_{H\psi_R} &= (H^\dagger i\overleftrightarrow{D}_\mu H)(\overline{\psi_R}\gamma^\mu\psi_R)\,,\\
\mathcal{O}_{H\psi_L}^{(1)} &= (H^\dagger i\overleftrightarrow{D}_\mu H)(\overline{\psi_L}\gamma^\mu\psi_L),\\
\mathcal{O}_{H\psi_L}^{(3)} &= (H^\dagger i\overleftrightarrow{D}_\mu^I H)(\overline{\psi_L}\gamma^\mu\sigma_I\psi_L),\\
\mathcal{O}_{Hud} &= (\tilde{H} i D_\mu H) (u \gamma^\mu d) + \text{h.c.}\,,
\end{align}
with $\psi_R = e,u,d$ and $\psi_L = l, q$.

Conversely, the renormalisation of $c_{H^4 D^4}^{(2)}$ driven by $\lambda$ can take it out of its positivity region. To show why, let us now consider the limit of negligible dimension-six terms (this is only for simplicity of the exposition) and also $c_{H^4 D^4}^{(2)\,\text{tree}}=0$. Upon deforming again the logarithm, we obtain:
\begin{align}
 \mathcal{A}(s) \sim \frac{g^2}{2} [\frac{3}{2\pi^2}(\lambda^\text{tree})^2\frac{\Lambda^4}{m^4} - \gamma_i \lambda^{\text{tree}} c_{H^4 D^4}^{(i)\,\text{tree}}\log{\frac{\Lambda^2}{m^2}}]\frac{s^2}{\Lambda^4}\,.
\end{align}
In the limit $m^2\to 0$, the first term dominates and therefore $\gamma_i c_{H^4 D^4}^{(i)\,\text{tree}}$ is not necessarily negative for arbitrary values of the Wilson coefficients. This conclusion still holds if dimension-six operators are not ignored, precisely because they do not contribute to $\mathcal{A}(s)$ at tree level, and because they fulfill Eq.~\eqref{eq:dim6ren}.

All these observations hold still in the presence of gauge couplings, although the proof is less straightforward. (For example, the massless gauge bosons induce poles at $s=0$ even at tree level.) Likewise, analyses analogous to the one we just did for $\phi_1\phi_2\to\phi_1\phi_2$ but applied to $\phi_1\phi_3\to\phi_1\phi_3$ and $\phi_1\phi_1\to\phi_1\phi_1$ reveal that the bounds in Eq.~\eqref{eq:otherbounds1}--\eqref{eq:otherbounds2} could be also violated at the loop level.\\

In summary, we can conclude that:
\begin{itemize}
 \item If effective interactions can arise at tree level, but either $c_{H^4 D^4}^{(2)}$ or the combination $c_{H^4 D^4}^{(1)}+c_{H^4 D^4}^{(2)}$ or $c_{H^4 D^4}^{(1)}+c_{H^4 D^4}^{(2)}+c_{H^4 D^4}^{(3)}$ vanishes accidentally at this order, then the constraints in Eqs.~\eqref{eq:firstconstraint}, \eqref{eq:otherbounds1} or~\eqref{eq:otherbounds2} can be broken, respectively.
 Interestingly, some popular extensions of the SM, \textit{e.g.} the addition of a scalar neutral singlet, fall in this category.
 \item If no operator can be generated at tree level, then all bounds in Eqs.~\eqref{eq:firstconstraint}--\eqref{eq:otherbounds2} are satisfied at one loop.
 \item The renormalisation of $c_{H^4 D^4}^{(j)}$  by pairs of dimension-six operators maintain those Wilson coefficients within their (tree-level) positivity region.
 \item The renormalisation of $c_{H^4 D^4}^{(j)}$ by relevant couplings (including mixing with other dimension-eight operators) can drive these Wilson coefficients out of their positivity region.
\end{itemize}
\section{One-loop matching of UV models}\label{sec:matching}
In the reminder of this paper, we show that our previous arguments, albeit somewhat heuristic, are in fact realised in minimal extensions of the SM. We refer to Appendix~\ref{app:details} for technical details.

First, let us extend the SM with a heavy scalar neutral singlet $\mathcal{S}$ of mass $M =\Lambda$, with interaction Lagrangian:
\begin{align}
 \mathcal{L}_\mathcal{S} = \kappa_{S}\mathcal{S} H^\dagger H\,.
\end{align}
This is obviously not the most generic Lagrangian, but it suffices to illustrate our point.

\begin{figure}[t]
 \includegraphics[width=\columnwidth]{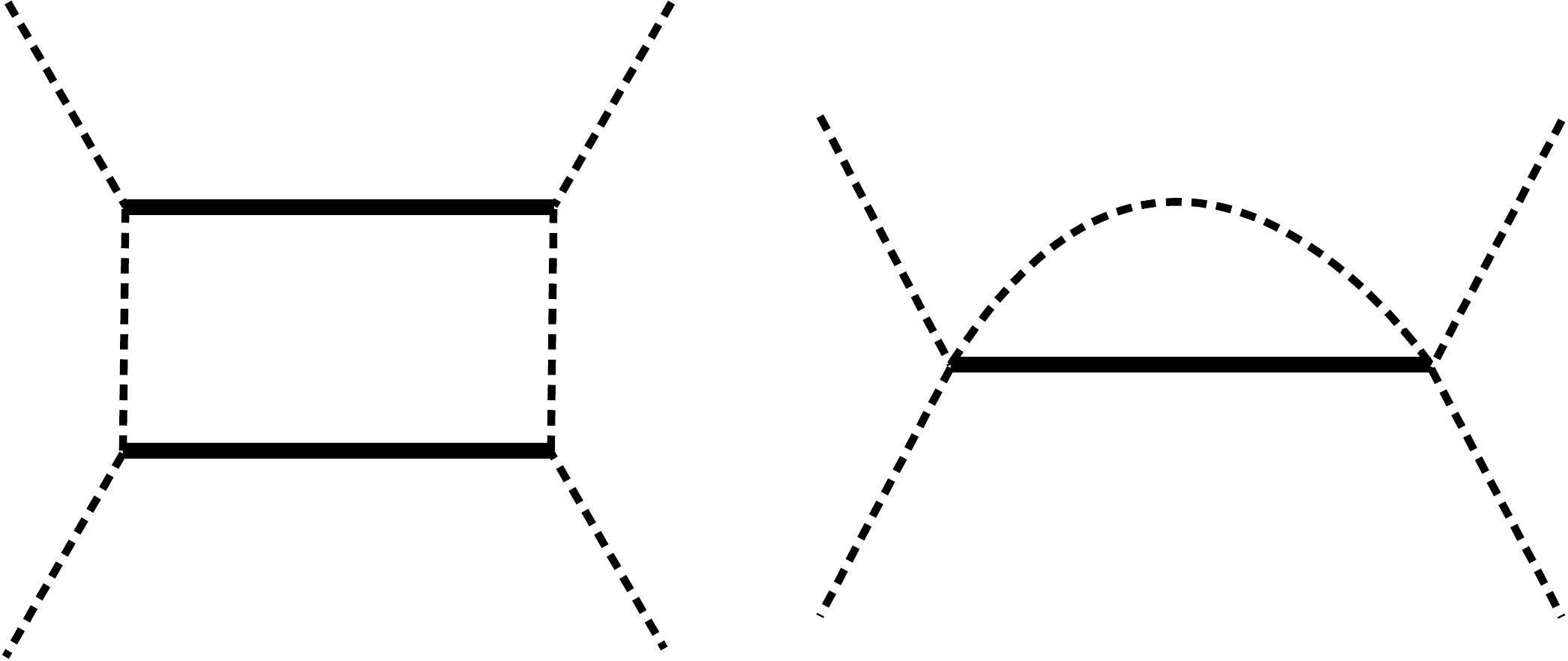}
 \caption{\it Example diagrams for integrating out a scalar singlet or triplet (left) or scalar quadruplets (right), giving rise to four-Higgs operators. The thick solid lines represent the heavy fields.}\label{fig:matching}
\end{figure}

At tree level, we obtain: 
\begin{equation}
 c_{H^4 D^4}^{(1)\,\text{tree}} = c_{H^4 D^4}^{(2)\,\text{tree}} = 0\,,\quad c_{H^4 D^4}^{(3)\,\text{tree}} = 2 \frac{\kappa_{\mathcal{S}}^2}{M^2}\,.
\end{equation}
At one loop (see Fig.~\ref{fig:matching}) and at the matching scale $\mu=M$, we get instead:
\begin{align}
 c_{H^4 D^4}^{(1)\,\text{loop}} &= -\frac{39}{144 \pi^2} \frac{\kappa_\mathcal{S}^4}{M^4}\,,\\
 c_{H^4 D^4}^{(2)\,\text{loop}} &= -\frac{39}{144 \pi^2} \frac{\kappa_\mathcal{S}^4}{M^4}\,,\\
 c_{H^4 D^4}^{(3)\,\text{loop}} &= -\frac{187}{720 \pi^2} \frac{\kappa_\mathcal{S}^4}{M^4}\,.
\end{align}
Therefore,
\begin{align}
 c_{H^4 D^4}^{(2)} &= -\frac{39}{144 \pi^2} \frac{\kappa_\mathcal{S}^4}{M^4} < 0\,,\\
 c_{H^4 D^4}^{(1)} + c_{H^4 D^4}^{(2)} &=-\frac{39}{72 \pi^2} \frac{\kappa_\mathcal{S}^4}{M^4} < 0\,,
\end{align}
and then both Eq.~\eqref{eq:firstconstraint} and Eq.~\eqref{eq:otherbounds1} are violated within this model.

Let us now consider the SM extended with a scalar real triplet $\Xi$ of mass $M$, too. The relevant Lagrangian is
\begin{equation}
 \mathcal{L}_{\Xi} = \kappa_{\Xi} H^\dagger \Xi^I \sigma^I H\,.
\end{equation}
When integrating $\Xi$ out up to one loop, we obtain:
\begin{align}
 \nonumber\\
 c_{H^4 D^4}^{(1)} &= 4\frac{\kappa_{\Xi}^2}{M^2}-\frac{107}{144\pi^2}\frac{\kappa_{\Xi}^4}{M^4}\,,\\
\nonumber\\
 c_{H^4 D^4}^{(2)} &= -\frac{61}{144 \pi^2}\frac{\kappa_{\Xi}^4}{ M^4}\,,\\
 c_{H^4 D^4}^{(3)} &= -2\frac{\kappa_{\Xi}^2}{M^2}-\frac{271}{720\pi^2}\frac{\kappa_{\Xi}^4}{M^4}\,.
\end{align}
The tree and loop contributions are manifest. Once again, $c_{H^4 D^2}^{(2)}<0$. Note also that, because there is no tree-level contribution to the four-Higgs operators from mixed diagrams involving heavy triplets and singlets, this bound is broken in any UV completion involving only this type of fields.

In all cases, dimension-six four-Higgs operators are also generated. Their explicit tree-level values can be found in Refs.~\cite{deBlas:2014mba,deBlas:2017xtg}.

Let us now turn our attention to three scalar extensions of the SM which \textit{do not} generate any four-Higgs operators at tree level (including those of dimension six). These involve adding a heavy doublet with $Y=1/2$ ($\varphi$) and adding heavy quadruplets with $Y=1/2$ ($\Theta_1$) and $\,\,\,\,\,\,\,\,\,Y=3/2$ ($\Theta_3$), respectively. We obtain:
\begin{align}
 c_{H^4 D^4}^{(1)} &= \frac{|\lambda_{\varphi}|^2}{24 \pi^2}\,,\\
 c_{H^4 D^4}^{(3)} &= \frac{|\lambda_{\varphi}|^2}{24 \pi^2}\,,\\
 c_{H^4 D^4}^{(3)} &= \frac{|\lambda_{\varphi}|^2}{6 \pi^2}\,;
\end{align}
as well as
\begin{align}
 c_{H^4 D^4}^{(1)} &= \frac{|\lambda_{\Theta_1}|^2}{9\pi^2}\,,\\
 c_{H^4 D^4}^{(2)} &= \frac{|\lambda_{\Theta_1}|^2}{36 \pi^2}\,,\\
 c_{H^4 D^4}^{(3)} &= -\frac{|\lambda_{\Theta_1}|^2}{18 \pi^2}\,;
\end{align}
and
\begin{align}
 c_{H^4 D^4}^{(1)} &= 0\,,\\
 c_{H^4 D^4}^{(2)} &= \frac{|\lambda_{\Theta_3}|^2}{4 \pi^2}\,,\\
 c_{H^4 D^4}^{(3)} &=0\,;
\end{align}
where $\lambda_\varphi$, $\lambda_{\Theta_1}$ and $\lambda_{\Theta_3}$ are the unique linear couplings between one heavy field and three $H$ bosons that can be written at the renormalisable level in each case; see Ref.~\cite{deBlas:2014mba}.

In all these cases, as we already anticipated, the conditions in Eqs.~\eqref{eq:firstconstraint}--\eqref{eq:otherbounds2} do hold.

\section{Renormalisation group evolution}\label{sec:running}
Let us now focus on the running of the Wilson coefficients $c_{H^4 D^4}^{(j)}$. The contribution triggered by pairs of dimension-six operators, computed in Ref.~\cite{Chala:2021pll}, reads:
\begin{widetext}
\begin{align}
 16\pi^2\beta_{H^4 D^4}^{(1)} &= \frac{8}{3}\bigg[-2 (c_{H^4 D^2}^{(1)})^2 - \frac{11}{8}(c_{H^4 D^2}^{(2)})^2 + 4c_{H^4 D^2}^{(1)}c_{H^4 D^2}^{(2)} \nonumber\\
 &\underline{+ 3 c_{H d}^2}\,\, \underline{\underline{+ c_{He}^2}}\,\, \underline{\underline{\underline{+ 2 (c_{Hl}^{(1)})^2}}} - 2 (c_{Hl}^{(3)})^2 \underline{\underline{\underline{\underline{+ 6 (c_{Hq}^{(1)})^2}}}} - 6 (c_{Hq}^{(3)})^2 + \underline{\underline{\underline{\underline{\underline{3 c_{H u}^2}}}}} - 3c_{Hud}^2\bigg]\,,\\[0.5cm]
 16\pi^2\beta_{H^4 D^4}^{(2)} &= \frac{8}{3}\bigg[-2 (c_{H^4 D^2}^{(1)})^2 - \frac{5}{8}(c_{H^4 D^2}^{(2)})^2 -2 c_{H^4 D^2}^{(1)}c_{H^4 D^2}^{(2)}\nonumber\\
&\underline{-3 c_{Hd}^2}\,\,\underline{\underline{-c_{He}^2}}\, \,\underline{\underline{\underline{-2 (c_{Hl}^{(1)})^2}}} -2 (c_{Hl}^{(3)})^2 \underline{\underline{\underline{\underline{- 6 (c_{Hq}^{(1)})^2}}}} - 6 (c_{Hq}^{(3)})^2 \underline{\underline{\underline{\underline{\underline{-3 c_{Hu}^2}}}}}\bigg]\,,\\[0.5cm]
 16\pi^2\beta_{H^4 D^4}^{(3)} &= \frac{8}{3}\bigg[-5 (c_{H^4 D^2}^{(1)})^2 + \frac{7}{8}(c_{H^4 D^2}^{(2)})^2 - 2 c_{H^4 D^2}^{(1)}c_{H^4 D^2}^{(2)} +4 (c_{Hl}^{(3)})^2 +12 (c_{Hq}^{(3)})^2 +3 c_{Hud}^2\bigg]\,.
\end{align}
\end{widetext}
(The Wilson coefficients of the fermionic operators are matrices in flavour space, so $c^2$ must be interpreted as the trace $\text{Tr}[c^\dagger c]$.)

It can be trivially seen, for example, that $\beta_{H^4 D^4}^{(2)} <0$, implying
\begin{equation}
c_{H^4 D^4}^{(2)}(\mu)\sim \beta_{H^4 D^4}^{(2)}\log{\frac{\mu}{\Lambda}} > 0\,,
\end{equation}
given that $\mu/\Lambda<1$ within the region of validity of the EFT.

Likewise, we have that $\beta_{H^4 D^4}^{(1)}+\beta_{H^4 D^4}^{(2)}<0$ as well as $\beta_{H^4 D^4}^{(1)}+\beta_{H^4 D^4}^{(2)}+\beta_{H^4 D^4}^{(3)}<0$, and therefore all bounds in Eqs.~\eqref{eq:firstconstraint}--\eqref{eq:otherbounds2} are respected by dimension-six quantum corrections at all scales. 

For illustration, in the equations above we have marked which positive (fermionic) coefficients in $\beta_{H^4 D^4}^{(1)}$ are canceled by $\beta_{H^4 D^4}^{(2)}$, forcing the second inequality even  though it could well be that $\beta_{H^4 D^4}^{(1)}>0$. Note also that $\beta_{H^4 D^4}^{(3)}$ can be positive in a plethora of cases, for example simply if $c_{Hl}^{(3)}$ is the only non-vanishing Wilson coefficient. The same holds for $\beta_{H^4 D^4}^{(1)}+\beta_{H^4 D^4}^{(3)}$ as well as for $\beta_{H^4 D^4}^{(2)}+\beta_{H^4 D^4}^{(3)}$. That is, the cancellations are in place precisely in those combinations given by the positivity bounds.

However, the mixing of the three $c_{H^4 D^4}^{(j)}$ due to renormalisable terms drive the former away from their positivity region. Indeed, upon a thorough computation based on the observations made in Appendix~\ref{app:details}, including the (in this case dominant) contribution from gauge couplings, we obtain that:
\begin{widetext}
\begin{align}
 16\pi^2\beta_{H^4 D^4}^{(1)} &= \frac{1}{6}\bigg[(30 c_{H^4 D^4}^{(1)} + 41 c_{H^4D^4}^{(2)} + 15 c_{H^4 D^4}^{(3)})g_2^2 - (16 c_{H^4 D^4}^{(1)}+7c_{H^4 D^4}^{(2)}+15c_{H^4 D^4}^{(3)}) g_1^2 \nonumber\\&+ 16(3 c_{H^4 D^4}^{(1)}+c_{H^4 D^4}^{(2)}+c_{H^4 D^4}^{(3)})\lambda \bigg] \,,\label{eq:dim8rges1}\\
 16\pi^2\beta_{H^4 D^4}^{(2)} &= \frac{1}{6}\bigg[(28 c_{H^4 D^4}^{(1)} + 43 c_{H^4D^4}^{(2)} + 15 c_{H^4 D^4}^{(3)})g_2^2 + (14 c_{H^4 D^4}^{(1)}+ 33 c_{H^4 D^4}^{(2)}+ 15 c_{H^4 D^4}^{(3)})g_1^2 \nonumber\\& + 16(c_{H^4 D^4}^{(1)}+3 c_{H^4 D^4}^{(2)}+c_{H^4 D^4}^{(3)})\lambda\bigg] \,,\label{eq:dim8rges2}\\
 16\pi^2\beta_{H^4 D^4}^{(3)} &= -\frac{1}{3}\bigg[(36 c_{H^4 D^4}^{(1)} + 29 c_{H^4D^4}^{(2)} + 42 c_{H^4 D^4}^{(3)})g_2^2 + (8c_{H^4 D^4}^{(1)}+2c_{H^4 D^4}^{(2)} + 9 c_{H^4 D^4}^{(3)})g_1^2 \nonumber\\&- 16(3 c_{H^4 D^4}^{(1)}+2 c_{H^4 D^4}^{(2)}+5 c_{H^4 D^4}^{(3)}) \lambda\bigg] \,.\label{eq:dim8rges3}
\end{align}
\end{widetext}
(We do not include fermionic dimension-eight operators, because they do not arise in the models in which we later use these expressions.)
It is clear that $\beta_{H^4 D^4}^{(2)}$ is not necessarily negative; likewise for the other positivity relations.

As a matter of example, let us assume that $\,\,\,\,\,\,\,c_{H^4 D^4}^{(2)}(\mu=M)=0$. Then, we have:
\begin{align}
 c_{H^4 D^4}^{(2)}(\mu) \sim &-\frac{1}{96 \pi^2} \bigg[(28 g_2^2 + 14g_1^2+16\lambda)c_{H^4 D^4}^{(1)}(M) \nonumber\\
 &+ (15 g_2^2 + 15 g_1^2 + 16\lambda) c_{H^4 D^4}^{(3)}(M)\bigg]\log{\frac{M}{\mu}}\,.
\end{align}  
If $c_{H^4 D^4}^{(1)}(\mu=M)\geq 0$ and $c_{H^4 D^4}^{(3)}(\mu=M)\geq 0$, as predicted for example in the neutral singlet scalar extension of the SM, then $c_{H^4 D^4}^{(2)}(\mu)$ is strictly negative, in conflict with Eq.~\eqref{eq:firstconstraint}. As we discussed above, this does not contradict the positivity of the forward scattering amplitude, as this is dominated by the running of the relevant couplings.

Similar violations of the positivity bounds occur in many other models for which some of the combinations of Wilson coefficients entering the inequalities in Eqs.~\eqref{eq:firstconstraint}--\eqref{eq:otherbounds2} vanish at tree level. To mention a few of the simplest ones:
\begin{align}
 \mathcal{S}\sim (1,1)_0 &\longmapsto c_{H^4 D^4}^{(1,2,3)} \sim (0,0,1)\label{eq:singlet}\,,\\
 \Xi\sim (1,3)_0 &\longmapsto c_{H^4 D^4}^{(1,2,3)} \sim (2,0,-1)\,,\\
 \mathcal{B}\sim (1,1)_0 &\longmapsto c_{H^4 D^4}^{(1,2,3)} \sim (-1,1,0)\,,\\
 \mathcal{B}_1\sim (1,1)_1 &\longmapsto c_{H^4 D^4}^{(1,2,3)} \sim (1,0,-1)\,,\\
 \mathcal{W}\sim (1,3)_0 &\longmapsto c_{H^4 D^4}^{(1,2,3)} \sim (1,1,-2)\label{eq:wvector}\,.
\end{align}
The first two fields are scalars, while the last three are vectors. The first numbers in parentheses and the subscript represent the $SU(3)_c\times SU(2)_L$ quantum numbers and the hypercharge, respectively. The last numbers in parentheses contain the ratios of the $c_{H^4 D^4}^{(1,2,3)}$ Wilson coefficients at tree level.

\begin{figure}[t]
 \includegraphics[width=1\columnwidth]{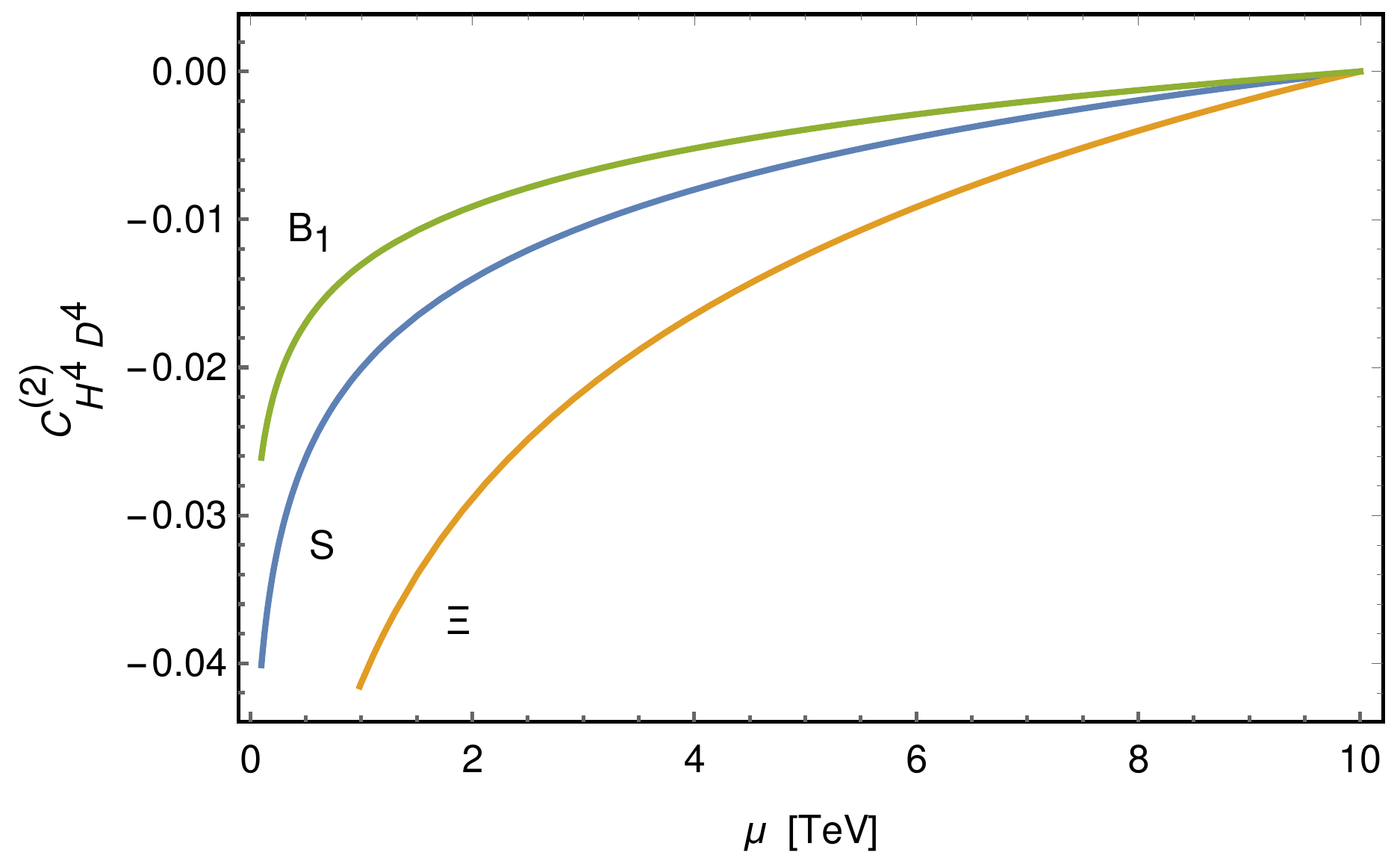}
\caption{\it Evolution of $c_{H^4 D^4}^{(2)}$ in different models in which $c_{H^4 D^4}^{(2)}(\mu=M)=0$ at the matching scale $M$; see the text for details.}\label{fig:running2}
\end{figure}

\begin{figure}[t]
 \includegraphics[width=1\columnwidth]{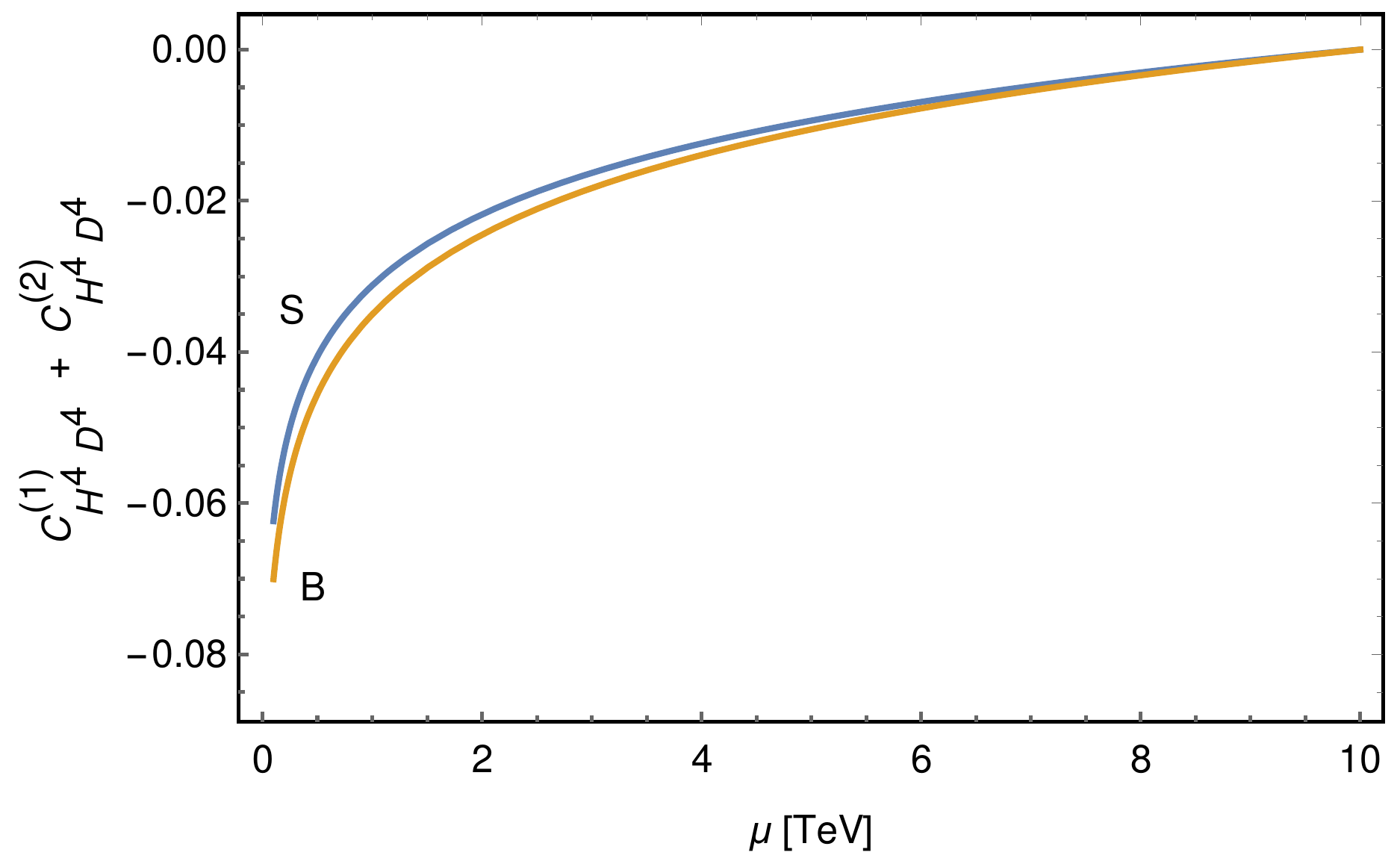}
\caption{\it Same as Fig.~\ref{fig:running2} but for $c_{H^4 D^4}^{(1)}+c_{H^4 D^4}^{(2)}$.}\label{fig:running12}
\end{figure}

\begin{figure}[t]
 \includegraphics[width=1\columnwidth]{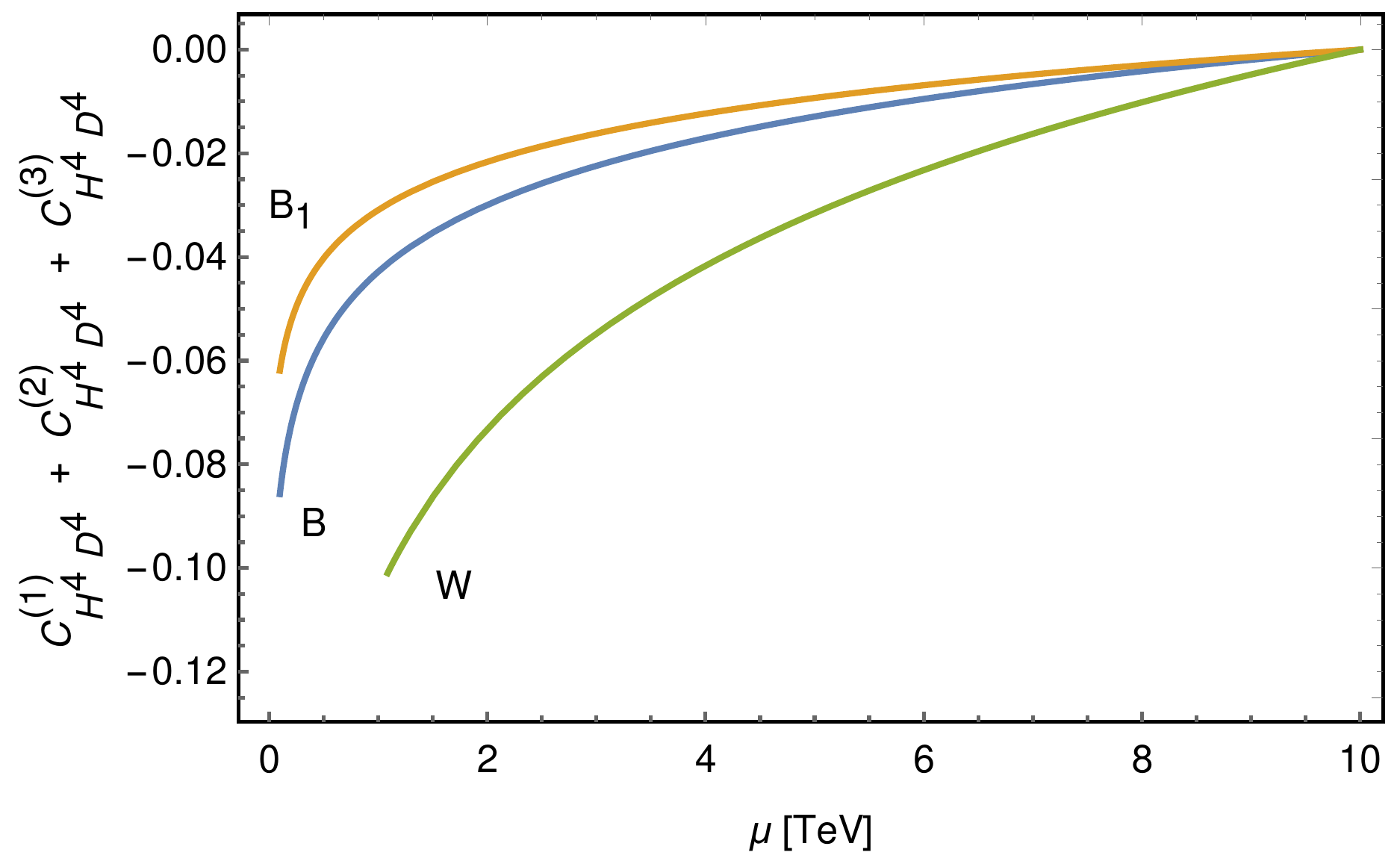}
\caption{\it Same as Fig.~\ref{fig:running2} but for $c_{H^4 D^4}^{(1)}+c_{H^4 D^4}^{(2)}+c_{H^4 D^4}^{(3)}$.}\label{fig:running123}
\end{figure}

We plot the evolution of $c_{H^4 D^4}^{(2)}$, $c_{H^4 D^4}^{(1)}+c_{H^4 D^4}^{(2)}$ and $c_{H^4 D^4}^{(1)}+c_{H^4 D^4}^{(2)}+c_{H^4 D^4}^{(3)}$ in Figs.~\ref{fig:running2}, \ref{fig:running12} and \ref{fig:running123}, respectively. For each model, we assume that the Wilson coefficients are fixed to the values given in Eqs.~\eqref{eq:singlet}--\eqref{eq:wvector} at the matching scale $M=10$ TeV. All curves include also the (sub-leading) contribution of dimension-six terms to the running. 

\section{Conclusions}\label{sec:conclusions}
We have argued that tree-level-derived positivity bounds on the Wilson coefficients $c_{H^4 D^4}^{(j)}$ of four-Higgs dimension-eight operators within the SMEFT do not necessarily hold at one loop.

First, they can be violated at the matching scale.
We have underpinned this statement with explicit calculations. In particular, we have computed the one-loop matching of some scalar extensions of the SM onto the SMEFT to dimension eight, demonstrating the violation of two of the three positivity bounds. 
And second, positivity bounds can be broken by their running triggered by renormalisable interactions such as the Higgs quartic or the gauge couplings. 

In this respect, it would be interesting to study modified scale-dependent constraints, relying on $s$-dependent integration contours such as the \textit{arcs} discussed in Ref.~\cite{Bellazzini:2020cot}; see also Section 11 of Ref.~\cite{Arkani-Hamed:2020blm}.~\footnote{During the completion of this work, Ref.~\cite{Baratella:2021guc} appeared on the arXiv. It shows that running effects from gravity \textit{do} preserve positivity.}

Conversely, we have shown that the renormalisation of $c_{H^4 D^4}^{(j)}$ driven by dimension-six terms does not break positivity. This implies strong constraints on the form of the corresponding anomalous dimensions.
In turn, this observation could provide new non-renormalisation results. For example, let us consider the renormalisation of a certain  Wilson coefficient $c_{8}$, subject to the bound $c_{8}>0$, by dimension-six terms $c_{6}^{(1)}$ and $c_{6}^{(2)}$:
\begin{align}
 \mu\frac{d c_{8}}{d\mu} = \gamma_{11} (c_{6}^{(1)})^2 + \gamma_{12} c_{6}^{(1)} c_{6}^{(2)} +\gamma_{22} (c_{6}^{(2)})^2\,.
\end{align}
Let us now assume that either $\gamma_{11}$ or $\gamma_{22}$ (or both) vanishes trivially (\textit{e.g.} because there are no Feynman diagrams, as it is the case in several examples shown in Ref.~\cite{Chala:2021pll}). Then, provided that $c_{6}^{(1,2)}$ can have either sign, $\gamma_{12}$ must be necessarily zero, or else $c_{8}$ could be negative for appropriate values of those terms. It would be worth exploring this avenue for understanding the surprising non-renormalisation of some dimension-eight bosonic operators~\cite{Chala:2021pll}, resulting from the exact cancellation of physical and redundant counterterms.

\section*{Acknowledgments} 
We thank P. Olgoso for help with \texttt{SuperTracer}. We thank A. D\'iaz-Carmona and A. Titov for useful discussions. This work has been supported by the SRA under grant number PID2019-106087GB-C21/C22 (10.13039/501100011033), and by
the Junta de Andaluc\'ia grants FQM 101, A-FQM-211-UGR18 and P18-FR-4314 (FEDER).
MC is also supported by the Spanish MINECO under the Ram\'on y Cajal programme.

\appendix

\section{Technical details}\label{app:details}
Let us start discussing the subtleties of the procedure for
matching the UV models in Section~\ref{sec:matching} onto the SMEFT to dimension eight.

We work off-shell, equating the hard region contribution of one-light-particle-irreducible Green functions in the UV, to the corresponding tree-level SMEFT amplitudes. We rely on \texttt{Matchmaker}~\cite{matchmaker} for this purpose. We have also cross-checked some results with the help of \texttt{SuperTracer}~\cite{Fuentes-Martin:2020udw}. 

We use the Green basis of dimension-eight four-Higgs interactions constructed in Ref.~\cite{Chala:2021pll}, which we reproduce in Table~\ref{tab:greenbasis} for completeness.
\begin{table}[t]
 \begin{tabular}{cc}
  \hline\hline
  $\mathcal{O}_{H^4 D^4}^{(1)}$ & $(D_\mu H^\dagger D_\nu H) (D^\nu H^\dagger D^\mu H)$\\[0.1cm]
  $\mathcal{O}_{H^4 D^4}^{(2)}$ & $(D_\mu H^\dagger D_\nu H) (D^\mu H^\dagger D^\nu H)$\\[0.1cm]
  $\mathcal{O}_{H^4 D^4}^{(3)}$ & $(D_\mu H^\dagger D^\mu H) (D^\nu H^\dagger D_\nu H)$\\[0.1cm]
  \hline
  $\mathcal{O}_{H^4 D^4}^{(4)}$ & $D_\mu H^\dagger D^\mu H (H^\dagger D^2 H+\text{h.c.})$\\[0.1cm]
  $\mathcal{O}_{H^4 D^4}^{(5)}$ & $D_\mu H^\dagger D^\mu (H^\dagger i D^2 H)+\text{h.c.})$\\[0.1cm]
  $\mathcal{O}_{H^4 D^4}^{(6)}$ & $(D_\mu H^\dagger H) (D^2 H^\dagger D_\mu H)+\text{h.c.}$\\[0.1cm]
  $\mathcal{O}_{H^4 D^4}^{(7)}$ & $(D_\mu H^\dagger H) (D^2 H^\dagger i D_\mu H)+\text{h.c.}$\\[0.1cm]
  $\mathcal{O}_{H^4 D^4}^{(8)}$ & $(D^2 H^\dagger H) (D^2 H^\dagger H)+\text{h.c.}$\\[0.1cm]
  $\mathcal{O}_{H^4 D^4}^{(9)}$ & $(D^2 H^\dagger H) (i D^2 H^\dagger H)+\text{h.c.}$\\[0.1cm]
  $\mathcal{O}_{H^4 D^4}^{(10)}$ & $(D^2 H^\dagger D^2 H)(H^\dagger H)$\\[0.1cm]
  $\mathcal{O}_{H^4 D^4}^{(11)}$ & $(H^\dagger D^2 H)(D^2 H^\dagger H)$\\[0.1cm]
  $\mathcal{O}_{H^4 D^4}^{(12)}$ & $(D_\mu H^\dagger  H)(D^\mu H^\dagger D^2 H)+\text{h.c.}$\\[0.1cm]
  $\mathcal{O}_{H^4 D^4}^{(13)}$ & $(D_\mu H^\dagger  H)(D^\mu H^\dagger i D^2 H)+\text{h.c.}$\\[0.1cm]
  \hline\hline
 \end{tabular}
 \caption{\it Dimension-eight four-Higgs SMEFT operators that are independent off-shell, namely they cannot be related to each other by integration by parts or using algebraic identities.}\label{tab:greenbasis}
\end{table}
On-shell, the $10$ operators below the horizontal line can be reduced to other SMEFT operators by means of field redefinitions (or equivalently in this case, by equations of motion~\cite{Criado:2018sdb}).
The interesting aspect of the basis above, though, is that none of those redundant operators contribute to any of the $\mathcal{O}_{H^4 D^4}^{(1,2,3)}$. The reason is that all them involve $D^2 H$, which by equations of motion, in the absence of the Higgs quadratic term, becomes~\cite{Barzinji:2018xvu}: 
%
%
\begin{align}\label{eq:eom}
D^2 H &= -2\lambda (H^\dagger H)H +\cdots\nonumber\\
&+ \frac{1}{\Lambda^2}\bigg\lbrace 3c_{H^6} (H^\dagger H)^2 H+ 2 c_{H^4 D^2}^{(1)} H\Box (H^\dagger H)\nonumber\\
&- c_{H^4 D^2}^{(2)}\bigg[ (D^\mu H) (H^\dagger \overleftrightarrow{D}_\mu H) + H \partial^\mu(H^\dagger D_\mu H)\bigg]\bigg\rbrace\,,
\end{align}
%
thus increasing the 
number of Higgs fields in the operators to six. The ellipses above represent fermionic interactions (which can be safely ignored), and $c_{H^6}$ stands for the Wilson coefficient of $(H^\dagger H)^3$.

Other operators which could \textit{a priori} contribute to $\mathcal{O}_{H^4 D^4}^{(1,2,3)}$ when the fields are on the mass shell include~\footnote{Operators involving gauge bosons can only contribute to four-Higgs operators proportionally to gauge couplings, which we have neglected in Section~\ref{sec:matching}.}:
\begin{align}
 \mathcal{O}_{H^4 D^2}^{(3)} &= (H^\dagger H) (D_\mu H)^\dagger (D^\mu H)\,,\\
 \mathcal{O}_{H^4 D^2}^{(4)} &= (H^\dagger H) D_\mu (H^\dagger i \overleftrightarrow{D}^\mu H)\,,\\
 %
 %
 \mathcal{O}_{H^2 D^4} &= D^2 H^\dagger D^2 H\,,\\
 \mathcal{O}_{H^2 D^6} &= D^2 H^\dagger D_\mu D_\nu D^\mu D^\nu H\,.
\end{align}
The first two operators undergo the same fate as $\mathcal{O}_{H^4 D^4}^{(4\cdots 13)}$; see Ref.~\cite{Chala:2021pll}. 
Here, we show that the two-Higgs operators $\mathcal{O}_{H^2 D^4}$ and $\mathcal{O}_{H^2 D^6}$ are also orthogonal to $\mathcal{O}_{H^4 D^4}^{(1,2,3)}$. This is trivial to see in the case of $\mathcal{O}_{H^2 D^4}$, which becomes $\mathcal{O}(H^6)$ by equations of motion~\footnote{
Let us note that two-Higgs operators can be only generated at one loop. Therefore, using the equations of motion to eliminate them from the action is still equivalent to performing field redefinitions (the difference is a to two-loop effect). 
}.

The operator $\mathcal{O}_{H^2 D^6}$ is slightly more subtle. Upon using the equations of motion, we obtain:
\begin{align}
 \mathcal{O}_{H^2 D^6} &= -2\lambda (H^\dagger H) H^\dagger D_\mu D^\nu D^\mu D^\nu H\nonumber\\
 &= 2\lambda \bigg[ 2\mathcal{O}_{H^4 D^4}^{(4)}+2\mathcal{O}_{H^4 D^4}^{(6)}+\mathcal{O}_{H^4 D^4}^{(8)}\nonumber\\
 &\,\,\,\,\,\,\,\,\,\,\,\,\,+\mathcal{O}_{H^4 D^4}^{(10)}
 +\mathcal{O}_{H^4 D^4}^{(11)}+\mathcal{O}_{H^4 D^4}^{(12)}\bigg]\,,
\end{align}
where in the second step we have only used integration by parts and algebraic identities.
Now, given that we end up with only redundant four-Higgs operators, and because none of them contribute to $\mathcal{O}_{H^4 D^4}^{(1,2,3)}$ on-shell as discussed previously, we conclude that $\mathcal{O}_{H^2 D^6}$ can be also disregarded.

Finally, modifications of the Higgs kinetic term can only affect each the $\mathcal{O}_{H^4 D^4}^{(1,2,3)}$ arising at tree level, because the Higgs wave-function can be only corrected at one loop.  We take this effect into account in the results in Section~\ref{sec:matching}, despite not introducing any qualitative change in the discussion. 

This suffices to demonstrate that, for the matching, the leading contribution to the Wilson coefficients $c_{H^4 D^4}^{(1,2,3)}$ can be fixed by simply computing the corresponding amplitudes $\phi_i\phi_j\to \phi_k\phi_l$ off-shell, projecting the result onto the basis of Table~\ref{tab:greenbasis}, and then just reading the first three coefficients.

The renormalisation of $c_{H^4D^4}^{(1,2,3)}$ through SM couplings becomes slightly more complicated, in particular because we have included the full gauge dependence.

We still work off-shell, but in this case we obviously restrict simply to (the divergent part of) one-particle-irreducible amplitudes. We use again \texttt{Matchmaker} for this matter, and we have fully cross-checked all results with \texttt{FeynArts}~\cite{Hahn:2000kx} and \texttt{FormCalc}~\cite{Hahn:1998yk}.

As commented in Section~\ref{sec:running}, we restrict to the mixing among $\mathcal{O}_{H^4 D^4}^{(1,2,3)}$, because no other relevant dimension-eight operator (in particular none with fermions) is generated at tree level in the models that we investigate therein. Now, because we are not neglecting gauge couplings, the following redundant operators involving gauge bosons must be also considered:\\
\begin{align}
 \mathcal{O}_{BH^2 D^4}^{(1)} &= \partial_\mu B^{\mu\nu} (D_\nu H^\dagger i D^2 H + \text{h.c.}) \,,\\
 \mathcal{O}_{BH^2 D^4}^{(2)} &=  \partial_\mu B^{\mu\nu} (D_\nu H^\dagger D^2 H + \text{h.c.})\,,\\
 \mathcal{O}_{BH^2 D^4}^{(3)} &= \partial_\mu B^{\mu\nu} (D_\rho D_\nu H^\dagger i D^\rho H+\text{h.c.}) \,,\\
 \mathcal{O}_{WH^2 D^4}^{(1)} &= D_\mu W^{\mu\nu I} (D_\nu H^\dagger i\sigma^I D^2 H + \text{h.c.})\,,\\[-0.3cm]
 & \nonumber\\[-0.2cm]
 \mathcal{O}_{WH^2 D^4}^{(2)} &=  D_\mu W^{\mu\nu I} (D_\nu H^\dagger \sigma^I D^2 H + \text{h.c.})\,,\\
&\nonumber\\[-0.3cm]
 \mathcal{O}_{WH^2 D^4}^{(3)} &= D_\mu W^{\mu\nu I} (D_\rho D_\nu H^\dagger i \sigma^I D^\rho H+\text{h.c.}) \,.
\end{align}
They expand the Green basis of dimension-eight operators with two Higgses, two derivatives and one field-strength. By construction, the corresponding Wilson coefficients are all real.

The only operators that contribute to four-Higgs interactions when moving on-shell are $\mathcal{O}_{BH^2 D^4}^{(3)}$ and $\mathcal{O}_{WH^2 D^4}^{(3)}$. Indeed, by using the equations of motion for $B$ and $W$:
\begin{align}
 \partial^\nu B_{\mu\nu} &= \frac{g_1}{2}H^\dagger i\overleftrightarrow{D}_\mu H+\cdots\,,\\
 D^\nu W^I_{\mu\nu} &= \frac{g_2}{2} H^\dagger i\overleftrightarrow{D}_\mu^I H+\cdots
\end{align}
(the ellipses encode again fermionic terms), we obtain, up to terms with two field strengths:
\begin{align}
 \mathcal{O}_{BH^2 D^4}^{(3)} &\sim \frac{g_1}{2} (H^\dagger \overleftrightarrow{D}_\nu H) (D_\rho D^\nu H^\dagger) D^\rho H+\text{h.c.}\nonumber\\
 &= g_1\bigg[-\mathcal{O}_{H^4 D^4}^{(1)}+ \mathcal{O}_{H^4 D^4}^{(2)} -\frac{1}{2}\mathcal{O}_{H^4 D^4}^{(6)} + \frac{1}{2}\mathcal{O}_{H^4 D^4}^{(12)}\bigg]\,,\\
 \mathcal{O}_{WH^2 D^4}^{(3)} &\sim \frac{g_2}{2} (H^\dagger \overleftrightarrow{D}^I_\nu H) (D_\rho D^\nu H^\dagger)\sigma^I D^\rho H+\text{h.c.}\nonumber\\
 &= g_2\bigg[\mathcal{O}_{H^4 D^4}^{(1)}+ \mathcal{O}_{H^4 D^4}^{(2)} - 2\mathcal{O}_{H^4 D^4}^{(3)} -2\mathcal{O}_{H^4 D^4}^{(4)} \nonumber\\
 &\,\,\,\,\,\,\,\,\,\,\,\,+ \frac{1}{2}\mathcal{O}_{H^4 D^4}^{(6)}+  \frac{1}{2}\mathcal{O}_{H^4 D^4}^{(12)}\bigg]\,.
\end{align}
Eqs.~\eqref{eq:dim8rges1}--\eqref{eq:dim8rges3} include these corrections, that we obtain from computing the off-shell amplitudes $\phi_i\to B\phi_i$ and $\phi_i\to W^3\phi_i$. Obviously, we also include the Higgs wavefunction renormalisation.

Finally, as for the matching, the running due to pairs of dimension-six terms can be computed by relying only on four-Higgs interactions, because there are no diagrams with only two external Higgs legs.

\bibliographystyle{apsrev4-1}
\bibliography{notes}

\begin{thebibliography}{33}%
\makeatletter
\providecommand \@ifxundefined [1]{%
 \@ifx{#1\undefined}
}%
\providecommand \@ifnum [1]{%
 \ifnum #1\expandafter \@firstoftwo
 \else \expandafter \@secondoftwo
 \fi
}%
\providecommand \@ifx [1]{%
 \ifx #1\expandafter \@firstoftwo
 \else \expandafter \@secondoftwo
 \fi
}%
\providecommand \natexlab [1]{#1}%
\providecommand \enquote  [1]{``#1''}%
\providecommand \bibnamefont  [1]{#1}%
\providecommand \bibfnamefont [1]{#1}%
\providecommand \citenamefont [1]{#1}%
\providecommand \href@noop [0]{\@secondoftwo}%
\providecommand \href [0]{\begingroup \@sanitize@url \@href}%
\providecommand \@href[1]{\@@startlink{#1}\@@href}%
\providecommand \@@href[1]{\endgroup#1\@@endlink}%
\providecommand \@sanitize@url [0]{\catcode `\\12\catcode `\$12\catcode
  `\&12\catcode `\#12\catcode `\^12\catcode `\_12\catcode `\%12\relax}%
\providecommand \@@startlink[1]{}%
\providecommand \@@endlink[0]{}%
\providecommand \url  [0]{\begingroup\@sanitize@url \@url }%
\providecommand \@url [1]{\endgroup\@href {#1}{\urlprefix }}%
\providecommand \urlprefix  [0]{URL }%
\providecommand \Eprint [0]{\href }%
\providecommand \doibase [0]{http://dx.doi.org/}%
\providecommand \selectlanguage [0]{\@gobble}%
\providecommand \bibinfo  [0]{\@secondoftwo}%
\providecommand \bibfield  [0]{\@secondoftwo}%
\providecommand \translation [1]{[#1]}%
\providecommand \BibitemOpen [0]{}%
\providecommand \bibitemStop [0]{}%
\providecommand \bibitemNoStop [0]{.\EOS\space}%
\providecommand \EOS [0]{\spacefactor3000\relax}%
\providecommand \BibitemShut  [1]{\csname bibitem#1\endcsname}%
\let\auto@bib@innerbib\@empty
\bibitem [{\citenamefont {Brivio}\ and\ \citenamefont
  {Trott}(2019)}]{Brivio:2017vri}%
  \BibitemOpen
  \bibfield  {author} {\bibinfo {author} {\bibfnamefont {I.}~\bibnamefont
  {Brivio}}\ and\ \bibinfo {author} {\bibfnamefont {M.}~\bibnamefont {Trott}},\
  }\href {\doibase 10.1016/j.physrep.2018.11.002} {\bibfield  {journal}
  {\bibinfo  {journal} {Phys. Rept.}\ }\textbf {\bibinfo {volume} {793}},\
  \bibinfo {pages} {1} (\bibinfo {year} {2019})},\ \Eprint
  {http://arxiv.org/abs/1706.08945} {arXiv:1706.08945 [hep-ph]} \BibitemShut
  {NoStop}%
\bibitem [{\citenamefont {de~Blas}\ \emph {et~al.}(2017)\citenamefont
  {de~Blas}, \citenamefont {Ciuchini}, \citenamefont {Franco}, \citenamefont
  {Mishima}, \citenamefont {Pierini}, \citenamefont {Reina},\ and\
  \citenamefont {Silvestrini}}]{deBlas:2016nqo}%
  \BibitemOpen
  \bibfield  {author} {\bibinfo {author} {\bibfnamefont {J.}~\bibnamefont
  {de~Blas}}, \bibinfo {author} {\bibfnamefont {M.}~\bibnamefont {Ciuchini}},
  \bibinfo {author} {\bibfnamefont {E.}~\bibnamefont {Franco}}, \bibinfo
  {author} {\bibfnamefont {S.}~\bibnamefont {Mishima}}, \bibinfo {author}
  {\bibfnamefont {M.}~\bibnamefont {Pierini}}, \bibinfo {author} {\bibfnamefont
  {L.}~\bibnamefont {Reina}}, \ and\ \bibinfo {author} {\bibfnamefont
  {L.}~\bibnamefont {Silvestrini}},\ }\href {\doibase 10.22323/1.282.0690}
  {\bibfield  {journal} {\bibinfo  {journal} {PoS}\ }\textbf {\bibinfo {volume}
  {ICHEP2016}},\ \bibinfo {pages} {690} (\bibinfo {year} {2017})},\ \Eprint
  {http://arxiv.org/abs/1611.05354} {arXiv:1611.05354 [hep-ph]} \BibitemShut
  {NoStop}%
\bibitem [{\citenamefont {Ellis}\ \emph {et~al.}(2021)\citenamefont {Ellis},
  \citenamefont {Madigan}, \citenamefont {Mimasu}, \citenamefont {Sanz},\ and\
  \citenamefont {You}}]{Ellis:2020unq}%
  \BibitemOpen
  \bibfield  {author} {\bibinfo {author} {\bibfnamefont {J.}~\bibnamefont
  {Ellis}}, \bibinfo {author} {\bibfnamefont {M.}~\bibnamefont {Madigan}},
  \bibinfo {author} {\bibfnamefont {K.}~\bibnamefont {Mimasu}}, \bibinfo
  {author} {\bibfnamefont {V.}~\bibnamefont {Sanz}}, \ and\ \bibinfo {author}
  {\bibfnamefont {T.}~\bibnamefont {You}},\ }\href {\doibase
  10.1007/JHEP04(2021)279} {\bibfield  {journal} {\bibinfo  {journal} {JHEP}\
  }\textbf {\bibinfo {volume} {04}},\ \bibinfo {pages} {279} (\bibinfo {year}
  {2021})},\ \Eprint {http://arxiv.org/abs/2012.02779} {arXiv:2012.02779
  [hep-ph]} \BibitemShut {NoStop}%
\bibitem [{\citenamefont {Ethier}\ \emph {et~al.}(2021)\citenamefont {Ethier},
  \citenamefont {Magni}, \citenamefont {Maltoni}, \citenamefont {Mantani},
  \citenamefont {Nocera}, \citenamefont {Rojo}, \citenamefont {Slade},
  \citenamefont {Vryonidou},\ and\ \citenamefont {Zhang}}]{Ethier:2021bye}%
  \BibitemOpen
  \bibfield  {author} {\bibinfo {author} {\bibfnamefont {J.~J.}\ \bibnamefont
  {Ethier}}, \bibinfo {author} {\bibfnamefont {G.}~\bibnamefont {Magni}},
  \bibinfo {author} {\bibfnamefont {F.}~\bibnamefont {Maltoni}}, \bibinfo
  {author} {\bibfnamefont {L.}~\bibnamefont {Mantani}}, \bibinfo {author}
  {\bibfnamefont {E.~R.}\ \bibnamefont {Nocera}}, \bibinfo {author}
  {\bibfnamefont {J.}~\bibnamefont {Rojo}}, \bibinfo {author} {\bibfnamefont
  {E.}~\bibnamefont {Slade}}, \bibinfo {author} {\bibfnamefont
  {E.}~\bibnamefont {Vryonidou}}, \ and\ \bibinfo {author} {\bibfnamefont
  {C.}~\bibnamefont {Zhang}},\ }\href@noop {} {\  (\bibinfo {year} {2021})},\
  \Eprint {http://arxiv.org/abs/2105.00006} {arXiv:2105.00006 [hep-ph]}
  \BibitemShut {NoStop}%
\bibitem [{\citenamefont {Zhang}\ and\ \citenamefont
  {Zhou}(2019)}]{Zhang:2018shp}%
  \BibitemOpen
  \bibfield  {author} {\bibinfo {author} {\bibfnamefont {C.}~\bibnamefont
  {Zhang}}\ and\ \bibinfo {author} {\bibfnamefont {S.-Y.}\ \bibnamefont
  {Zhou}},\ }\href {\doibase 10.1103/PhysRevD.100.095003} {\bibfield  {journal}
  {\bibinfo  {journal} {Phys. Rev. D}\ }\textbf {\bibinfo {volume} {100}},\
  \bibinfo {pages} {095003} (\bibinfo {year} {2019})},\ \Eprint
  {http://arxiv.org/abs/1808.00010} {arXiv:1808.00010 [hep-ph]} \BibitemShut
  {NoStop}%
\bibitem [{\citenamefont {Bi}\ \emph {et~al.}(2019)\citenamefont {Bi},
  \citenamefont {Zhang},\ and\ \citenamefont {Zhou}}]{Bi:2019phv}%
  \BibitemOpen
  \bibfield  {author} {\bibinfo {author} {\bibfnamefont {Q.}~\bibnamefont
  {Bi}}, \bibinfo {author} {\bibfnamefont {C.}~\bibnamefont {Zhang}}, \ and\
  \bibinfo {author} {\bibfnamefont {S.-Y.}\ \bibnamefont {Zhou}},\ }\href
  {\doibase 10.1007/JHEP06(2019)137} {\bibfield  {journal} {\bibinfo  {journal}
  {JHEP}\ }\textbf {\bibinfo {volume} {06}},\ \bibinfo {pages} {137} (\bibinfo
  {year} {2019})},\ \Eprint {http://arxiv.org/abs/1902.08977} {arXiv:1902.08977
  [hep-ph]} \BibitemShut {NoStop}%
\bibitem [{\citenamefont {Remmen}\ and\ \citenamefont
  {Rodd}(2019)}]{Remmen:2019cyz}%
  \BibitemOpen
  \bibfield  {author} {\bibinfo {author} {\bibfnamefont {G.~N.}\ \bibnamefont
  {Remmen}}\ and\ \bibinfo {author} {\bibfnamefont {N.~L.}\ \bibnamefont
  {Rodd}},\ }\href {\doibase 10.1007/JHEP12(2019)032} {\bibfield  {journal}
  {\bibinfo  {journal} {JHEP}\ }\textbf {\bibinfo {volume} {12}},\ \bibinfo
  {pages} {032} (\bibinfo {year} {2019})},\ \Eprint
  {http://arxiv.org/abs/1908.09845} {arXiv:1908.09845 [hep-ph]} \BibitemShut
  {NoStop}%
\bibitem [{\citenamefont {Gu}\ \emph {et~al.}(2020)\citenamefont {Gu},
  \citenamefont {Wang},\ and\ \citenamefont {Zhang}}]{Gu:2020ldn}%
  \BibitemOpen
  \bibfield  {author} {\bibinfo {author} {\bibfnamefont {J.}~\bibnamefont
  {Gu}}, \bibinfo {author} {\bibfnamefont {L.-T.}\ \bibnamefont {Wang}}, \ and\
  \bibinfo {author} {\bibfnamefont {C.}~\bibnamefont {Zhang}},\ }\href@noop {}
  {\  (\bibinfo {year} {2020})},\ \Eprint {http://arxiv.org/abs/2011.03055}
  {arXiv:2011.03055 [hep-ph]} \BibitemShut {NoStop}%
\bibitem [{\citenamefont {Bonnefoy}\ \emph {et~al.}(2021)\citenamefont
  {Bonnefoy}, \citenamefont {Gendy},\ and\ \citenamefont
  {Grojean}}]{Bonnefoy:2020yee}%
  \BibitemOpen
  \bibfield  {author} {\bibinfo {author} {\bibfnamefont {Q.}~\bibnamefont
  {Bonnefoy}}, \bibinfo {author} {\bibfnamefont {E.}~\bibnamefont {Gendy}}, \
  and\ \bibinfo {author} {\bibfnamefont {C.}~\bibnamefont {Grojean}},\ }\href
  {\doibase 10.1007/JHEP04(2021)115} {\bibfield  {journal} {\bibinfo  {journal}
  {JHEP}\ }\textbf {\bibinfo {volume} {04}},\ \bibinfo {pages} {115} (\bibinfo
  {year} {2021})},\ \Eprint {http://arxiv.org/abs/2011.12855} {arXiv:2011.12855
  [hep-ph]} \BibitemShut {NoStop}%
\bibitem [{\citenamefont {Remmen}\ and\ \citenamefont
  {Rodd}(2020{\natexlab{a}})}]{Remmen:2020vts}%
  \BibitemOpen
  \bibfield  {author} {\bibinfo {author} {\bibfnamefont {G.~N.}\ \bibnamefont
  {Remmen}}\ and\ \bibinfo {author} {\bibfnamefont {N.~L.}\ \bibnamefont
  {Rodd}},\ }\href {\doibase 10.1103/PhysRevLett.125.081601} {\bibfield
  {journal} {\bibinfo  {journal} {Phys. Rev. Lett.}\ }\textbf {\bibinfo
  {volume} {125}},\ \bibinfo {pages} {081601} (\bibinfo {year}
  {2020}{\natexlab{a}})},\ \Eprint {http://arxiv.org/abs/2004.02885}
  {arXiv:2004.02885 [hep-ph]} \BibitemShut {NoStop}%
\bibitem [{\citenamefont {Zhang}\ and\ \citenamefont
  {Zhou}(2020)}]{Zhang:2020jyn}%
  \BibitemOpen
  \bibfield  {author} {\bibinfo {author} {\bibfnamefont {C.}~\bibnamefont
  {Zhang}}\ and\ \bibinfo {author} {\bibfnamefont {S.-Y.}\ \bibnamefont
  {Zhou}},\ }\href {\doibase 10.1103/PhysRevLett.125.201601} {\bibfield
  {journal} {\bibinfo  {journal} {Phys. Rev. Lett.}\ }\textbf {\bibinfo
  {volume} {125}},\ \bibinfo {pages} {201601} (\bibinfo {year} {2020})},\
  \Eprint {http://arxiv.org/abs/2005.03047} {arXiv:2005.03047 [hep-ph]}
  \BibitemShut {NoStop}%
\bibitem [{\citenamefont {Gu}\ and\ \citenamefont {Wang}(2021)}]{Gu:2020thj}%
  \BibitemOpen
  \bibfield  {author} {\bibinfo {author} {\bibfnamefont {J.}~\bibnamefont
  {Gu}}\ and\ \bibinfo {author} {\bibfnamefont {L.-T.}\ \bibnamefont {Wang}},\
  }\href {\doibase 10.1007/JHEP03(2021)149} {\bibfield  {journal} {\bibinfo
  {journal} {JHEP}\ }\textbf {\bibinfo {volume} {03}},\ \bibinfo {pages} {149}
  (\bibinfo {year} {2021})},\ \Eprint {http://arxiv.org/abs/2008.07551}
  {arXiv:2008.07551 [hep-ph]} \BibitemShut {NoStop}%
\bibitem [{\citenamefont {Fuks}\ \emph {et~al.}(2021)\citenamefont {Fuks},
  \citenamefont {Liu}, \citenamefont {Zhang},\ and\ \citenamefont
  {Zhou}}]{Fuks:2020ujk}%
  \BibitemOpen
  \bibfield  {author} {\bibinfo {author} {\bibfnamefont {B.}~\bibnamefont
  {Fuks}}, \bibinfo {author} {\bibfnamefont {Y.}~\bibnamefont {Liu}}, \bibinfo
  {author} {\bibfnamefont {C.}~\bibnamefont {Zhang}}, \ and\ \bibinfo {author}
  {\bibfnamefont {S.-Y.}\ \bibnamefont {Zhou}},\ }\href {\doibase
  10.1088/1674-1137/abcd8c} {\bibfield  {journal} {\bibinfo  {journal} {Chin.
  Phys. C}\ }\textbf {\bibinfo {volume} {45}},\ \bibinfo {pages} {023108}
  (\bibinfo {year} {2021})},\ \Eprint {http://arxiv.org/abs/2009.02212}
  {arXiv:2009.02212 [hep-ph]} \BibitemShut {NoStop}%
\bibitem [{\citenamefont {Yamashita}\ \emph {et~al.}(2021)\citenamefont
  {Yamashita}, \citenamefont {Zhang},\ and\ \citenamefont
  {Zhou}}]{Yamashita:2020gtt}%
  \BibitemOpen
  \bibfield  {author} {\bibinfo {author} {\bibfnamefont {K.}~\bibnamefont
  {Yamashita}}, \bibinfo {author} {\bibfnamefont {C.}~\bibnamefont {Zhang}}, \
  and\ \bibinfo {author} {\bibfnamefont {S.-Y.}\ \bibnamefont {Zhou}},\ }\href
  {\doibase 10.1007/JHEP01(2021)095} {\bibfield  {journal} {\bibinfo  {journal}
  {JHEP}\ }\textbf {\bibinfo {volume} {01}},\ \bibinfo {pages} {095} (\bibinfo
  {year} {2021})},\ \Eprint {http://arxiv.org/abs/2009.04490} {arXiv:2009.04490
  [hep-ph]} \BibitemShut {NoStop}%
\bibitem [{\citenamefont {Remmen}\ and\ \citenamefont
  {Rodd}(2020{\natexlab{b}})}]{Remmen:2020uze}%
  \BibitemOpen
  \bibfield  {author} {\bibinfo {author} {\bibfnamefont {G.~N.}\ \bibnamefont
  {Remmen}}\ and\ \bibinfo {author} {\bibfnamefont {N.~L.}\ \bibnamefont
  {Rodd}},\ }\href@noop {} {\  (\bibinfo {year} {2020}{\natexlab{b}})},\
  \Eprint {http://arxiv.org/abs/2010.04723} {arXiv:2010.04723 [hep-ph]}
  \BibitemShut {NoStop}%
\bibitem [{\citenamefont {Adams}\ \emph {et~al.}(2006)\citenamefont {Adams},
  \citenamefont {Arkani-Hamed}, \citenamefont {Dubovsky}, \citenamefont
  {Nicolis},\ and\ \citenamefont {Rattazzi}}]{Adams:2006sv}%
  \BibitemOpen
  \bibfield  {author} {\bibinfo {author} {\bibfnamefont {A.}~\bibnamefont
  {Adams}}, \bibinfo {author} {\bibfnamefont {N.}~\bibnamefont {Arkani-Hamed}},
  \bibinfo {author} {\bibfnamefont {S.}~\bibnamefont {Dubovsky}}, \bibinfo
  {author} {\bibfnamefont {A.}~\bibnamefont {Nicolis}}, \ and\ \bibinfo
  {author} {\bibfnamefont {R.}~\bibnamefont {Rattazzi}},\ }\href {\doibase
  10.1088/1126-6708/2006/10/014} {\bibfield  {journal} {\bibinfo  {journal}
  {JHEP}\ }\textbf {\bibinfo {volume} {10}},\ \bibinfo {pages} {014} (\bibinfo
  {year} {2006})},\ \Eprint {http://arxiv.org/abs/hep-th/0602178}
  {arXiv:hep-th/0602178} \BibitemShut {NoStop}%
\bibitem [{\citenamefont {Bellazzini}\ \emph {et~al.}(2021)\citenamefont
  {Bellazzini}, \citenamefont {Elias~Mir\'o}, \citenamefont {Rattazzi},
  \citenamefont {Riembau},\ and\ \citenamefont {Riva}}]{Bellazzini:2020cot}%
  \BibitemOpen
  \bibfield  {author} {\bibinfo {author} {\bibfnamefont {B.}~\bibnamefont
  {Bellazzini}}, \bibinfo {author} {\bibfnamefont {J.}~\bibnamefont
  {Elias~Mir\'o}}, \bibinfo {author} {\bibfnamefont {R.}~\bibnamefont
  {Rattazzi}}, \bibinfo {author} {\bibfnamefont {M.}~\bibnamefont {Riembau}}, \
  and\ \bibinfo {author} {\bibfnamefont {F.}~\bibnamefont {Riva}},\ }\href
  {\doibase 10.1103/PhysRevD.104.036006} {\bibfield  {journal} {\bibinfo
  {journal} {Phys. Rev. D}\ }\textbf {\bibinfo {volume} {104}},\ \bibinfo
  {pages} {036006} (\bibinfo {year} {2021})},\ \Eprint
  {http://arxiv.org/abs/2011.00037} {arXiv:2011.00037 [hep-th]} \BibitemShut
  {NoStop}%
\bibitem [{\citenamefont {Arkani-Hamed}\ \emph {et~al.}(2021)\citenamefont
  {Arkani-Hamed}, \citenamefont {Huang},\ and\ \citenamefont
  {Huang}}]{Arkani-Hamed:2020blm}%
  \BibitemOpen
  \bibfield  {author} {\bibinfo {author} {\bibfnamefont {N.}~\bibnamefont
  {Arkani-Hamed}}, \bibinfo {author} {\bibfnamefont {T.-C.}\ \bibnamefont
  {Huang}}, \ and\ \bibinfo {author} {\bibfnamefont {Y.-T.}\ \bibnamefont
  {Huang}},\ }\href {\doibase 10.1007/JHEP05(2021)259} {\bibfield  {journal}
  {\bibinfo  {journal} {JHEP}\ }\textbf {\bibinfo {volume} {05}},\ \bibinfo
  {pages} {259} (\bibinfo {year} {2021})},\ \Eprint
  {http://arxiv.org/abs/2012.15849} {arXiv:2012.15849 [hep-th]} \BibitemShut
  {NoStop}%
\bibitem [{\citenamefont {Craig}\ \emph {et~al.}(2020)\citenamefont {Craig},
  \citenamefont {Jiang}, \citenamefont {Li},\ and\ \citenamefont
  {Sutherland}}]{Craig:2019wmo}%
  \BibitemOpen
  \bibfield  {author} {\bibinfo {author} {\bibfnamefont {N.}~\bibnamefont
  {Craig}}, \bibinfo {author} {\bibfnamefont {M.}~\bibnamefont {Jiang}},
  \bibinfo {author} {\bibfnamefont {Y.-Y.}\ \bibnamefont {Li}}, \ and\ \bibinfo
  {author} {\bibfnamefont {D.}~\bibnamefont {Sutherland}},\ }\href {\doibase
  10.1007/JHEP08(2020)086} {\bibfield  {journal} {\bibinfo  {journal} {JHEP}\
  }\textbf {\bibinfo {volume} {08}},\ \bibinfo {pages} {086} (\bibinfo {year}
  {2020})},\ \Eprint {http://arxiv.org/abs/2001.00017} {arXiv:2001.00017
  [hep-ph]} \BibitemShut {NoStop}%
\bibitem [{\citenamefont {Grzadkowski}\ \emph {et~al.}(2010)\citenamefont
  {Grzadkowski}, \citenamefont {Iskrzynski}, \citenamefont {Misiak},\ and\
  \citenamefont {Rosiek}}]{Grzadkowski:2010es}%
  \BibitemOpen
  \bibfield  {author} {\bibinfo {author} {\bibfnamefont {B.}~\bibnamefont
  {Grzadkowski}}, \bibinfo {author} {\bibfnamefont {M.}~\bibnamefont
  {Iskrzynski}}, \bibinfo {author} {\bibfnamefont {M.}~\bibnamefont {Misiak}},
  \ and\ \bibinfo {author} {\bibfnamefont {J.}~\bibnamefont {Rosiek}},\ }\href
  {\doibase 10.1007/JHEP10(2010)085} {\bibfield  {journal} {\bibinfo  {journal}
  {JHEP}\ }\textbf {\bibinfo {volume} {10}},\ \bibinfo {pages} {085} (\bibinfo
  {year} {2010})},\ \Eprint {http://arxiv.org/abs/1008.4884} {arXiv:1008.4884
  [hep-ph]} \BibitemShut {NoStop}%
\bibitem [{\citenamefont {Murphy}(2020)}]{Murphy:2020rsh}%
  \BibitemOpen
  \bibfield  {author} {\bibinfo {author} {\bibfnamefont {C.~W.}\ \bibnamefont
  {Murphy}},\ }\href {\doibase 10.1007/JHEP10(2020)174} {\bibfield  {journal}
  {\bibinfo  {journal} {JHEP}\ }\textbf {\bibinfo {volume} {10}},\ \bibinfo
  {pages} {174} (\bibinfo {year} {2020})},\ \Eprint
  {http://arxiv.org/abs/2005.00059} {arXiv:2005.00059 [hep-ph]} \BibitemShut
  {NoStop}%
\bibitem [{\citenamefont {Froissart}(1961)}]{Froissart:1961ux}%
  \BibitemOpen
  \bibfield  {author} {\bibinfo {author} {\bibfnamefont {M.}~\bibnamefont
  {Froissart}},\ }\href {\doibase 10.1103/PhysRev.123.1053} {\bibfield
  {journal} {\bibinfo  {journal} {Phys. Rev.}\ }\textbf {\bibinfo {volume}
  {123}},\ \bibinfo {pages} {1053} (\bibinfo {year} {1961})}\BibitemShut
  {NoStop}%
\bibitem [{\citenamefont {Martin}(1963)}]{Martin:1962rt}%
  \BibitemOpen
  \bibfield  {author} {\bibinfo {author} {\bibfnamefont {A.}~\bibnamefont
  {Martin}},\ }\href {\doibase 10.1103/PhysRev.129.1432} {\bibfield  {journal}
  {\bibinfo  {journal} {Phys. Rev.}\ }\textbf {\bibinfo {volume} {129}},\
  \bibinfo {pages} {1432} (\bibinfo {year} {1963})}\BibitemShut {NoStop}%
\bibitem [{\citenamefont {Chala}\ \emph {et~al.}(2021)\citenamefont {Chala},
  \citenamefont {Guedes}, \citenamefont {Ramos},\ and\ \citenamefont
  {Santiago}}]{Chala:2021pll}%
  \BibitemOpen
  \bibfield  {author} {\bibinfo {author} {\bibfnamefont {M.}~\bibnamefont
  {Chala}}, \bibinfo {author} {\bibfnamefont {G.}~\bibnamefont {Guedes}},
  \bibinfo {author} {\bibfnamefont {M.}~\bibnamefont {Ramos}}, \ and\ \bibinfo
  {author} {\bibfnamefont {J.}~\bibnamefont {Santiago}},\ }\href {\doibase
  10.21468/SciPostPhys.11.3.065} {\bibfield  {journal} {\bibinfo  {journal}
  {SciPost Phys.}\ }\textbf {\bibinfo {volume} {11}},\ \bibinfo {pages} {065}
  (\bibinfo {year} {2021})},\ \Eprint {http://arxiv.org/abs/2106.05291}
  {arXiv:2106.05291 [hep-ph]} \BibitemShut {NoStop}%
\bibitem [{\citenamefont {de~Blas}\ \emph {et~al.}(2015)\citenamefont
  {de~Blas}, \citenamefont {Chala}, \citenamefont {Perez-Victoria},\ and\
  \citenamefont {Santiago}}]{deBlas:2014mba}%
  \BibitemOpen
  \bibfield  {author} {\bibinfo {author} {\bibfnamefont {J.}~\bibnamefont
  {de~Blas}}, \bibinfo {author} {\bibfnamefont {M.}~\bibnamefont {Chala}},
  \bibinfo {author} {\bibfnamefont {M.}~\bibnamefont {Perez-Victoria}}, \ and\
  \bibinfo {author} {\bibfnamefont {J.}~\bibnamefont {Santiago}},\ }\href
  {\doibase 10.1007/JHEP04(2015)078} {\bibfield  {journal} {\bibinfo  {journal}
  {JHEP}\ }\textbf {\bibinfo {volume} {04}},\ \bibinfo {pages} {078} (\bibinfo
  {year} {2015})},\ \Eprint {http://arxiv.org/abs/1412.8480} {arXiv:1412.8480
  [hep-ph]} \BibitemShut {NoStop}%
\bibitem [{\citenamefont {de~Blas}\ \emph {et~al.}(2018)\citenamefont
  {de~Blas}, \citenamefont {Criado}, \citenamefont {Perez-Victoria},\ and\
  \citenamefont {Santiago}}]{deBlas:2017xtg}%
  \BibitemOpen
  \bibfield  {author} {\bibinfo {author} {\bibfnamefont {J.}~\bibnamefont
  {de~Blas}}, \bibinfo {author} {\bibfnamefont {J.~C.}\ \bibnamefont {Criado}},
  \bibinfo {author} {\bibfnamefont {M.}~\bibnamefont {Perez-Victoria}}, \ and\
  \bibinfo {author} {\bibfnamefont {J.}~\bibnamefont {Santiago}},\ }\href
  {\doibase 10.1007/JHEP03(2018)109} {\bibfield  {journal} {\bibinfo  {journal}
  {JHEP}\ }\textbf {\bibinfo {volume} {03}},\ \bibinfo {pages} {109} (\bibinfo
  {year} {2018})},\ \Eprint {http://arxiv.org/abs/1711.10391} {arXiv:1711.10391
  [hep-ph]} \BibitemShut {NoStop}%
\bibitem [{\citenamefont {Baratella}\ \emph {et~al.}(2021)\citenamefont
  {Baratella}, \citenamefont {Haslehner}, \citenamefont {Ruhdorfer},
  \citenamefont {Serra},\ and\ \citenamefont {Weiler}}]{Baratella:2021guc}%
  \BibitemOpen
  \bibfield  {author} {\bibinfo {author} {\bibfnamefont {P.}~\bibnamefont
  {Baratella}}, \bibinfo {author} {\bibfnamefont {D.}~\bibnamefont
  {Haslehner}}, \bibinfo {author} {\bibfnamefont {M.}~\bibnamefont
  {Ruhdorfer}}, \bibinfo {author} {\bibfnamefont {J.}~\bibnamefont {Serra}}, \
  and\ \bibinfo {author} {\bibfnamefont {A.}~\bibnamefont {Weiler}},\
  }\href@noop {} {\  (\bibinfo {year} {2021})},\ \Eprint
  {http://arxiv.org/abs/2109.06191} {arXiv:2109.06191 [hep-th]} \BibitemShut
  {NoStop}%
\bibitem [{\citenamefont {Carmona}\ \emph {et~al.}()\citenamefont {Carmona},
  \citenamefont {Lazopoulos}, \citenamefont {Olgoso},\ and\ \citenamefont
  {Santiago}}]{matchmaker}%
  \BibitemOpen
  \bibfield  {author} {\bibinfo {author} {\bibfnamefont {A.}~\bibnamefont
  {Carmona}}, \bibinfo {author} {\bibfnamefont {A.}~\bibnamefont {Lazopoulos}},
  \bibinfo {author} {\bibfnamefont {P.}~\bibnamefont {Olgoso}}, \ and\ \bibinfo
  {author} {\bibfnamefont {J.}~\bibnamefont {Santiago}},\ }\href@noop {}
  {\bibinfo  {journal} {(to appear)}\ }\BibitemShut {NoStop}%
\bibitem [{\citenamefont {Fuentes-Martin}\ \emph {et~al.}(2021)\citenamefont
  {Fuentes-Martin}, \citenamefont {K\"onig}, \citenamefont {Pag\`es},
  \citenamefont {Thomsen},\ and\ \citenamefont
  {Wilsch}}]{Fuentes-Martin:2020udw}%
  \BibitemOpen
\bibfield  {journal} {  }\bibfield  {author} {\bibinfo {author} {\bibfnamefont
  {J.}~\bibnamefont {Fuentes-Martin}}, \bibinfo {author} {\bibfnamefont
  {M.}~\bibnamefont {K\"onig}}, \bibinfo {author} {\bibfnamefont
  {J.}~\bibnamefont {Pag\`es}}, \bibinfo {author} {\bibfnamefont {A.~E.}\
  \bibnamefont {Thomsen}}, \ and\ \bibinfo {author} {\bibfnamefont
  {F.}~\bibnamefont {Wilsch}},\ }\href {\doibase 10.1007/JHEP04(2021)281}
  {\bibfield  {journal} {\bibinfo  {journal} {JHEP}\ }\textbf {\bibinfo
  {volume} {04}},\ \bibinfo {pages} {281} (\bibinfo {year} {2021})},\ \Eprint
  {http://arxiv.org/abs/2012.08506} {arXiv:2012.08506 [hep-ph]} \BibitemShut
  {NoStop}%
\bibitem [{\citenamefont {Criado}\ and\ \citenamefont
  {P\'erez-Victoria}(2019)}]{Criado:2018sdb}%
  \BibitemOpen
  \bibfield  {author} {\bibinfo {author} {\bibfnamefont {J.~C.}\ \bibnamefont
  {Criado}}\ and\ \bibinfo {author} {\bibfnamefont {M.}~\bibnamefont
  {P\'erez-Victoria}},\ }\href {\doibase 10.1007/JHEP03(2019)038} {\bibfield
  {journal} {\bibinfo  {journal} {JHEP}\ }\textbf {\bibinfo {volume} {03}},\
  \bibinfo {pages} {038} (\bibinfo {year} {2019})},\ \Eprint
  {http://arxiv.org/abs/1811.09413} {arXiv:1811.09413 [hep-ph]} \BibitemShut
  {NoStop}%
\bibitem [{\citenamefont {Barzinji}\ \emph {et~al.}(2018)\citenamefont
  {Barzinji}, \citenamefont {Trott},\ and\ \citenamefont
  {Vasudevan}}]{Barzinji:2018xvu}%
  \BibitemOpen
  \bibfield  {author} {\bibinfo {author} {\bibfnamefont {A.}~\bibnamefont
  {Barzinji}}, \bibinfo {author} {\bibfnamefont {M.}~\bibnamefont {Trott}}, \
  and\ \bibinfo {author} {\bibfnamefont {A.}~\bibnamefont {Vasudevan}},\ }\href
  {\doibase 10.1103/PhysRevD.98.116005} {\bibfield  {journal} {\bibinfo
  {journal} {Phys. Rev. D}\ }\textbf {\bibinfo {volume} {98}},\ \bibinfo
  {pages} {116005} (\bibinfo {year} {2018})},\ \Eprint
  {http://arxiv.org/abs/1806.06354} {arXiv:1806.06354 [hep-ph]} \BibitemShut
  {NoStop}%
\bibitem [{\citenamefont {Hahn}(2001)}]{Hahn:2000kx}%
  \BibitemOpen
  \bibfield  {author} {\bibinfo {author} {\bibfnamefont {T.}~\bibnamefont
  {Hahn}},\ }\href {\doibase 10.1016/S0010-4655(01)00290-9} {\bibfield
  {journal} {\bibinfo  {journal} {Comput. Phys. Commun.}\ }\textbf {\bibinfo
  {volume} {140}},\ \bibinfo {pages} {418} (\bibinfo {year} {2001})},\ \Eprint
  {http://arxiv.org/abs/hep-ph/0012260} {arXiv:hep-ph/0012260} \BibitemShut
  {NoStop}%
\bibitem [{\citenamefont {Hahn}\ and\ \citenamefont
  {Perez-Victoria}(1999)}]{Hahn:1998yk}%
  \BibitemOpen
  \bibfield  {author} {\bibinfo {author} {\bibfnamefont {T.}~\bibnamefont
  {Hahn}}\ and\ \bibinfo {author} {\bibfnamefont {M.}~\bibnamefont
  {Perez-Victoria}},\ }\href {\doibase 10.1016/S0010-4655(98)00173-8}
  {\bibfield  {journal} {\bibinfo  {journal} {Comput. Phys. Commun.}\ }\textbf
  {\bibinfo {volume} {118}},\ \bibinfo {pages} {153} (\bibinfo {year}
  {1999})},\ \Eprint {http://arxiv.org/abs/hep-ph/9807565}
  {arXiv:hep-ph/9807565} \BibitemShut {NoStop}%
\end{thebibliography}%

\end{document}